\documentclass[useAMS,usenatbib]{mn2e}
\usepackage{amsmath}
\usepackage{amssymb}
\usepackage{amsfonts}
\usepackage{xcolor}
\usepackage[draft]{graphicx}

\bibpunct{(}{)}{;}{a}{,}{,}

\makeatletter
\def\ref@jnl#1{{\rmfamily#1}}%
\newcommand\aj{\ref@jnl{AJ}}%
\newcommand\araa{\ref@jnl{ARA\&A}}%
\newcommand\apj{\ref@jnl{ApJ}}%
\newcommand\apjl{\ref@jnl{ApJ}}%
\newcommand\apjs{\ref@jnl{ApJS}}%
\newcommand\ao{\ref@jnl{Appl.~Opt.}}%
\newcommand\apss{\ref@jnl{Ap\&SS}}%
\newcommand\aap{\ref@jnl{A\&A}}%
\newcommand\aapr{\ref@jnl{A\&A~Rev.}}%
\newcommand\aaps{\ref@jnl{A\&AS}}%
\newcommand\azh{\ref@jnl{AZh}}%
\newcommand\baas{\ref@jnl{BAAS}}%
\newcommand\jrasc{\ref@jnl{JRASC}}%
\newcommand\memras{\ref@jnl{MmRAS}}%
\newcommand\mnras{\ref@jnl{MNRAS}}%
\newcommand\pra{\ref@jnl{Phys.~Rev.~A}}%
\newcommand\prb{\ref@jnl{Phys.~Rev.~B}}%
\newcommand\prc{\ref@jnl{Phys.~Rev.~C}}%
\newcommand\prd{\ref@jnl{Phys.~Rev.~D}}%
\newcommand\pre{\ref@jnl{Phys.~Rev.~E}}%
\newcommand\prl{\ref@jnl{Phys.~Rev.~Lett.}}%
\newcommand\pasp{\ref@jnl{PASP}}%
\newcommand\pasj{\ref@jnl{PASJ}}%
\newcommand\qjras{\ref@jnl{QJRAS}}%
\newcommand\skytel{\ref@jnl{S\&T}}%
\newcommand\solphys{\ref@jnl{Sol.~Phys.}}%
\newcommand\sovast{\ref@jnl{Soviet~Ast.}}%
\newcommand\ssr{\ref@jnl{Space~Sci.~Rev.}}%
\newcommand\zap{\ref@jnl{ZAp}}%
\newcommand\nat{\ref@jnl{Nature}}%
\newcommand\iaucirc{\ref@jnl{IAU~Circ.}}%
\newcommand\aplett{\ref@jnl{Astrophys.~Lett.}}%
\newcommand\apspr{\ref@jnl{Astrophys.~Space~Phys.~Res.}}%
\newcommand\bain{\ref@jnl{Bull.~Astron.~Inst.~Netherlands}}%
\newcommand\fcp{\ref@jnl{Fund.~Cosmic~Phys.}}%
\newcommand\gca{\ref@jnl{Geochim.~Cosmochim.~Acta}}%
\newcommand\grl{\ref@jnl{Geophys.~Res.~Lett.}}%
\newcommand\jcp{\ref@jnl{J.~Chem.~Phys.}}%
\newcommand\jgr{\ref@jnl{J.~Geophys.~Res.}}%
\newcommand\jqsrt{\ref@jnl{J.~Quant.~Spec.~Radiat.~Transf.}}%
\newcommand\memsai{\ref@jnl{Mem.~Soc.~Astron.~Italiana}}%
\newcommand\nphysa{\ref@jnl{Nucl.~Phys.~A}}%
\newcommand\physrep{\ref@jnl{Phys.~Rep.}}%
\newcommand\physscr{\ref@jnl{Phys.~Scr}}%
\newcommand\planss{\ref@jnl{Planet.~Space~Sci.}}%
\newcommand\procspie{\ref@jnl{Proc.~SPIE}}%

\long\def\comment#1{}

\def\ba{\begin{eqnarray}}
\def\ea{\end{eqnarray}}
\def\be{\begin{equation}}
\def\ee{\end{equation}}

\def\mR{\mathbfss{R}}

\def\bdx{\bmath{x }}

\def\bdw{\bmath{w }}
\def\bdk{\bmath{k }}

\def\bda{\bmath{a }}

\def\bdn{\bmath{n }}

\def\bdelta{{\delta }}

\newcommand{\tR}{{\mR}}

\def\nobs{N_{\rm obs}}

\title[SZ component separation from heterogeneous data exploitation]{Reconstruction of high-resolution SZ maps from heterogeneous datasets using needlets}
\author[Mathieu Remazeilles, Nabila Aghanim, Marian Douspis]{Mathieu Remazeilles\thanks{E-mail: mathieu.remazeilles@ias.u-psud.fr}, Nabila Aghanim, Marian Douspis\\
Institut d'Astrophysique Spatiale, Universit\'e Paris-Sud 11 \& CNRS (UMR 8617), B\^at. 121, 91405 Orsay Cedex, France}

\begin{document}


\pagerange{\pageref{firstpage}--\pageref{lastpage}} \pubyear{2012}

\maketitle

\label{firstpage}

\begin{abstract}
The aim of this work is to propose a joint exploitation of heterogeneous datasets from high-resolution/few-channel experiments and low-resolution/many-channel experiments by using a multiscale needlet Internal Linear Combination (ILC), in order to optimize the thermal Sunyaev-Zeldovich (SZ) effect reconstruction at high resolution. We highlight that needlet ILC is a powerful and tunable component separation method which can easily deal with multiple experiments with various specifications. Such a multiscale analysis renders possible the joint exploitation of high-resolution and low-resolution data, by performing for each needlet scale a combination of some specific channels, either from one dataset or both datasets, selected for their relevance to the angular scale considered, thus allowing to simultaneously extract high resolution SZ signal from compact clusters and remove Galactic foreground contamination at large scales.          
\end{abstract}


\section{Introduction}

Multifrequency observations of the microwave sky are a mixture of various diffuse and compact components \citep{1995SSRv...74...37B,1999NewA....4..443B}: the Cosmic Microwave Background (CMB) emission, the SZ effect from galaxy clusters, the foreground emission from the Galactic interstellar medium (ISM), the Cosmic Infrared Background (CIB) emission, the emission from compact extra-galactic radio sources, the instrumental noise.
Each of these components has a distinctive frequency signature (which can be known or not), a distinctive spatial distribution on the celestial sphere, and a distinctive spectral distribution on the angular scales (power spectrum).
The separation of the sky components in such multifrequency observations of the CMB is an important part of the processing
and {analysis} of such observational data. 
The thermal Sunyaev-Zeldovich (TSZ) effect from galaxy clusters is a spectral distortion of the CMB black body radiation due to inverse Compton scattering of the CMB photons off hot electrons contained in the intra-cluster gas \citep{1972CoASP...4..173S}. It is responsible for secondary temperature anisotropies, which introduce an excess of power in the primordial CMB temperature anisotropies at small angular scales. This spectral distortion is independent of the cosmological redshift so that the measure of the TSZ effect is a powerful and unique tool to detect new galaxy clusters at any redshift \citep{1999PhR...310...97B, 2002ARA&A..40..643C}. 
The separation and the extraction of the TSZ effect from the other sky emissions  is made possible by a multifrequency coherent analysis because of the prior knowledge of the frequency signature of the SZ effect. The main limitation in detecting SZ clusters comes from the achievable resolution of the instrument and the level of contamination by the other sky emissions and the instrumental noise. Various component separation methods have been developed to extract the thermal SZ signal from the observed frequency maps of a single experiment. Some of them require a prior assumption on the template SZ profile such as the matched filtering methods in \citet{1996MNRAS.279..545H}, \citet{2002ApJ...580..610H,2002MNRAS.336.1057H}, and \citet{2006A&A...459..341M}. Other methods are blind such as the methods based on statistical independence (ICA in \citet{1999ISPL....6..145H,2002MNRAS.334...53M}, Spectral Matching ICA in \citet{2003MNRAS.346.1089D,2008arXiv0803.1814C}, GMCA in \citet{2008StMet...5..307B} which also benefits from the sparsity of the components). Some methods are parametric and require physical modeling of some components (MEM in \citet{1998MNRAS.300....1H}, Commander in \citet{2008ApJ...676...10E}). Non-parametric methods include needlet Internal Linear Combination (NILC) in \citet{2008A&A...491..597L} and in \citet{2009A&A...493..835D}, MILCA in \citet{2010arXiv1007.1149H}, multidimensional NILC in \citet{2011MNRAS.410.2481R} and \citet{2011MNRAS.418..467R}). 
The ILC, which has been used first on the data of the WMAP mission to reconstruct the CMB emission \citep{2003ApJS..148...97B}, is a particularly simple component separation method which does not assume any particular parametrization for foreground emission, and for the other emissions considered as contaminants. The frequency scaling vector of the component of interest to extract is the only parameter needed for implementing ILC. For all the component separation methods, the problem is to find the best compromise between the most accurate unbiased estimation of the component of interest and the best minimization of the residuals from the other sky emissions.

In this work we address the problem of jointly exploiting the datasets of multiple  experiments with various specifications for optimizing the reconstruction of the thermal SZ effect at high resolution. 
No single instrument provides the best measurement of the SZ effect.
Space-borne instruments, such as Planck for instance \citep{planck2011-5.1b, 2011arXiv1112.5595P, 2012arXiv1205.3376P}, can observe the whole sky in the millimetre through many frequencies (nine channels from $30$ GHz to $857$ GHz). The large number of frequency channels covered by such space-borne instruments enables component separation methods, such as ILC, to improve the minimization of various contaminations (Galactic foregrounds, CMB, instrumental noise) in the reconstructed SZ map. However, the instrumental resolution of Planck for instance is limited to $5$ arcmin, therefore the extraction of SZ signal from clusters of arcminute angular size, by using such dataset only, remains unachievable. Conversely, higher resolution ground-based telescopes, such as the Atacama Cosmology Telescope (ACT) \citep{2011ApJ...731..100M} and the South Pole Telescope (SPT) \citep{2011ApJ...738..139W, 2012arXiv1203.5775R} for instance, are designed to measure the SZ temperature anisotropies up to arcminute angular scales but the presence of the atmosphere limits the frequency coverage of those experiments. The few number of frequency channels observed by such ground-based telescopes ($148$ GHz, $218$ GHz, and $277$ GHz in ACT) limits the ability of these instruments to remove the Galactic foreground contamination in the reconstruction of the SZ signal of nearby clusters. 
Here we propose to perform SZ component separation from the joint exploitation of heterogeneous sets of maps (both maps from many-channel/lower-resolution instrument and maps from few-channel/higher-resolution instrument) by using needlet ILC as a \emph{multiscale}, or \emph{multiresolution}, approach for combining multiple instrument datasets. 

The paper is organised as follows. In Sect. \ref{sec:method} we describe the needlet ILC method on patches of the sky and show how the needlets can be exploited to combine multiple instrument datasets with various resolutions for optimizing the SZ component separation. The results are presented on simulations of the sky in Sect. \ref{sec:results} where we apply the method jointly on two heterogeneous instrument datasets. We discuss the robustness of the approach in Sect. \ref{sec:discussion} and conclude in Sect. \ref{sec:conclusion}.



\section{Aggregating datasets with needlet ILC}\label{sec:method}

\subsection{Derivation of ILC}

All the available observation maps ($\nobs$ maps, indexed by $i$ for each frequency channel) can be written as
\ba
x_i(p) &=& a_i s(p) + n_i (p),
\ea
which can be recast in a vector form as
\ba
\bdx(p) & = & \bda s(p) + \bdn (p).
\ea
Here $p$ may denote either a pixel in a direct-space representation, or an $(\ell, m)$ index (resp. $\bdk=(k_1,k_2)$) in a harmonic space (resp. Fourier space) representation, or even a wavelet domain $(j;q)$ in a wavelet frame decomposition, where $j$ denotes the scale of the wavelet and $q$ the pixel. 
The vector $\bdx(p)$ collects the $\nobs$ observed frequency maps, $s(p)$ is the unknown thermal SZ template map that we would like to reconstruct, and $\bda$ is the known frequency scaling vector of the SZ effect. All the other sky emissions (CMB and foregrounds) and the instrumental noise are collected in a single nuisance term $\bdn(p)$. The ILC estimate of the SZ map, $\widehat{s} = \sum_i w_i x_i(p)$, is a linear combination of the observed maps of minimum variance under the constraint of offering unit response to the SZ component, so that the ILC weights $\bdw$ are solution to the following constrained minimum variance problem:
\ba\label{qqq:opt}
\left\{ \begin{tabular}{ll}
 $\mathrm{var} \left(\widehat{s} (p)\right)$ & ~$\mathrm{minimum,}$ \\ 
$\sum_i~ w_i a_i = 1$ & \\
\end{tabular} 
\right.
\ea
The first condition in Eq. (\ref{qqq:opt}) guarantees the minimum contamination by the background noise (residual sky components and instrumental noise) whereas the second condition guarantees the unbiased reconstruction of the SZ template. Note that the quality of the reconstruction relies on the accurate knowledge of the SZ frequency scaling $\bda$. In presence of calibration errors, there is no guarantee that the SZ component is conserved \citep{2010MNRAS.401.1602D}. Straightforward algebra using Lagrange multiplier \citep{2004ApJ...612..633E} leads to the following ILC estimate as the solution to  Eq. (\ref{qqq:opt}):
\ba\label{qqq:ilc}
\widehat{s}(p) = \bdw^T \bdx(p) = {\bda^T\widehat{\tR}^{-1}\over \bda^T\widehat{\tR}^{-1}\bda}\bdx(p),
\ea
where $\widehat{\tR}_{ii'} = {1\over N_{\mathrm{p}}}\sum_{p} x_i(p)x_{i'}(p)$ is the empirical covariance matrix of the observations. By construction, we have $\widehat{s}(p) = s(p) + \bdw^T \bdn(p)$, with the residual nuisance term minimized. The ILC relies on the component of interest to be uncorrelated with the contaminants (i.e. $\langle s(p)n_i (p)\rangle = 0$, for each channel). We discuss the case where this assumption can fail in Sect. \ref{subsec:bias} when applying ILC on small patches of the sky.


Let us now discuss the drawbacks of using either pixel-based or multipole-based ILC filtering in the context of exploiting heterogeneous datasets from multiple instruments with different specifications. 


The ILC filtering in pixel space \citep{2004ApJ...612..633E}, which is ultra-local in direct space, suffers from nonlocality in harmonic (or Fourier)  space by conjugation. 
Such a pixel-based ILC thus requires to combine the same set of channel-maps for all the angular scales considered. 
This is not optimal when the channel-maps have not the same angular resolution since it requires that the fixed set of maps be degraded to a common resolution 
which is compelled to be the lowest resolution of the set of channel-maps, so that the ILC reconstructed map is also compelled to be produced at this lowest resolution and the small scale information is lost.

Consequently, pixel-based ILC is either a resolution-limited method or a channel-limited method: suppose that we would like to reconstruct an SZ map at high resolution, say one arcminute, we would have to combine only the few channel-maps having a resolution which is higher than or equal to one arcminute. A reduced number of channels for SZ ILC reconstruction is not optimal for removing the contamination by other sky components because the number of sky components becomes larger than the number of channels, making the inverse problem suboptimal (also see the results in Sect. \ref{sec:results}). Conversely, if we choose to combine a larger number of channel-maps, for instance by including in the pixel-based ILC combination lower resolution channel-maps from another instrument, in order to better remove the background noise, then the ILC output SZ map would be restricted to have the lowest original resolution of the combined channel-maps. 


As an alternative, the component separation can in principle be performed in harmonic space \citep{2003PhRvD..68l3523T, 2008arXiv0803.1394K}.
%
A multipole-based ILC can mix maps with different resolutions, unlike a pixel-based method, by combining all the channel-maps at low $\ell$ and only the high resolution channel-maps at higher $\ell$. A harmonic space ILC (ultra-local in $\ell$) is a fortiori nonlocal in pixel space so that such an approach is better suited for all-sky analysis. Therefore, a multipole-based ILC can not easily handle data from a partial sky coverage. Ground-based experiments cover finite size sky area. The advantage of such partial sky surveys is that one is left with analyzing zones of the sky which are less contaminated by other sky emissions such as Galactic foregrounds. Here we are interested on performing component separation on patches of the sky thus we cannot use a harmonic space approach. 

\subsection{Needlet ILC: multiresolution}\label{subsec:nilc}

For the purposes of a joint analysis of heterogeneous datasets (many channels-lower resolution/few channels-higher resolution), we propose to use a wavelet-based component separation called Needlet ILC \citep{2009A&A...493..835D} because of the ability of a wavelet-based approach to combine data with different resolutions, unlike a pixel-based approach, and to handle patches of the sky, unlike a harmonic space approach. The needlets are a particular type of wavelets where the widths of the scale windows are freely tunable in harmonic space \citep{guilloux:fay:cardoso:2008}. The ability of wavelets for combining multi-instrument datasets has been highlighted in the case of CMB power spectrum reconstruction in \citet{2008PhRvD..78h3013F}.    

Instead of requiring strict locality either in pixel or in multipole space, we can perform a needlet-based component separation by decomposing the maps on a needlet frame.  The formulation of the needlets in the context of CMB data analysis can be found in \citet{2008MNRAS.383..539M} and \citet{guilloux:fay:cardoso:2008}. The needlet decomposition provides localization of the ILC filter both in pixel and in harmonic space, making the ILC filtering able to adapt to the local conditions of contamination (local Galactic contamination, small scale noise contamination). Wavelet localization has also been adopted by \citet{2012arXiv1206.1773B} in the GMCA method, based on sparsity, for optimizing component separation. The property of localization of the needlets both in scale and in space is particularly meaningful for an SZ dedicated component separation because the signal is non-uniformly distributed both in space (on the celestial sphere) and in scale (relevant at small scales). Furthermore, Needlet ILC (NILC) is a multiresolution approach allowing for combining channel-maps with different resolutions by using localization in Fourier or harmonic space (needlet bands). Moreover, such a wavelet approach  allows us to combine patches of the sky by using localization in pixel space.

Let us define a collection of window functions (Fig. \ref{Fig:bandsvsbeams}) $h_{k}^{(j)}$ in the Fourier domain, indexed by $j$, such that over the useful range of spatial frequency $k$, we have
\ba\label{qqq:bandconst}
\sum_j \left[h^{(j)}_k\right]^2 =1.
\ea 
In the following we will perform component separation on patches of the sky smaller than $100$ square degrees so that we adopt a flat-sky approximation allowing for using Fast Fourier transforms (FFT) instead of spherical harmonic transforms. Maps of needlet coefficients $x^{(j)}(p)$ \emph{at a given scale $(j)$} are obtained, for the observed maps $\bdx(p)$, by inverse FFT
of the associated map Fourier coefficients $\bdx_{k}$ filtered by the spectral windows $h^{(j)}_k$:
\ba\label{qqq:ana}
\bdx^{(j)}(p) = \sum_{k_1=0}^{N_k -1}\sum_{k_2=0}^{N_k -1}  h^{(j)}_k \bdx_{k} \exp{\left[i2\pi \left({k_1\over N_k} p_1 + {k_2\over N_k} p_2\right)\right]}
\ea
where $p=(p_1,p_2)$, $k=\sqrt{k_1^2+k_2^2}$ and $N_k$ is the number of resolution elements on each side of the two-dimensional FFT patch considered.
The Needlet ILC estimate at the scale $(j)$ is given by
\ba
\widehat{s}^{(j)}(p) = \bdw^T \bdx^{(j)}(p) = {\bda^T(\widehat{\tR}^{(j)}(p))^{-1}\over \bda^T(\widehat{\tR}^{(j)}(p))^{-1}\bda}\bdx^{(j)}(p),
\ea
where the local covariance matrix of the needlet coefficients of the observations is \ba\label{qqq:localcovar}\widehat{\tR}_{ab}^{(j)}(p) = {1\over N_{\mathrm{p}}} \sum_{p'\in \mathcal{D}_p} x^{(j)}_a(p')x^{(j)}_b(p').\ea 
The pixel domain $\mathcal{D}_p$, on which the local covariance is computed, is defined from the smoothing of the product map $x^{(j)}_a(p)x^{(j)}_b(p)$ with a symmetric Gaussian window in pixel space. The complete reconstruction of the SZ map $\widehat{s}(p)$ can then be obtained by simply coadding the reconstructed SZ ``maps per scale'' $\widehat{z}^{(j)}(p)$, 
\ba\label{qqq:coadd}
\widehat{s}(p) = \sum_j \widehat{z}^{(j)}(p),
\ea
where the SZ ``maps per scale'' themselves are obtained from the following scheme
\ba\label{qqq:scheme}
 \widehat{s}^{(j)}(p) \stackrel{\rm FFT}{\longrightarrow} h^{(j)}_k \widehat{s}_{k} \stackrel{\times}{\longrightarrow} \left(h^{(j)}_k\right)^2 \widehat{s}_{k} \stackrel{\rm FFT^{-1}}{\longrightarrow} \widehat{z}^{(j)}(p).
\ea
The constraint on the needlet scale-bands in Eq. (\ref{qqq:bandconst}) guarantees the total conservation of the SZ power by the processing in Eqs. (\ref{qqq:coadd}) and (\ref{qqq:scheme}).

\subsection{Example of needlets for heterogeneous datasets}\label{subsec:example}

The needlet ILC component separation has already been dedicated to all-sky TSZ reconstruction \citep{2008A&A...491..597L}. Here we develop needlet ILC on patches of sky 
 to focus on zones of the sky centered on particular clusters. Our main purpose is to use needlets on patches as a multiresolution tool for combining multiple heterogeneous experiments by ILC to optimize the TSZ component separation.

Let us illustrate how such a multiscale approach to component separation is relevant to the joint exploitation of heterogeneous data with different resolutions. In this regard we consider two experiments noted \textsc{exp\oldstylenums{1}} and \textsc{exp\oldstylenums{2}}, the former having a lower resolution $\theta_1$ than the latter having a beam $\theta_2 < \theta_1$, but a larger number of frequency channels $N_1 > N_2$. For the sake of simplicity we assume that the $N_1$ (resp. $N_2$) channel-maps of \textsc{exp\oldstylenums{1}} (resp. \textsc{exp\oldstylenums{2}}) have the same resolution $\theta_1$ (resp. $\theta_2$). The goal is to benefit both from the maximum number of channels $N_1+N_2$ and from the highest resolution $\theta_2$ for performing NILC component separation, both requirements being needed to guarantee a reconstructed TSZ signal with  the lowest sky residual contamination and with the highest resolution.  

\begin{figure}
  \begin{center}
    \includegraphics[width=8cm]{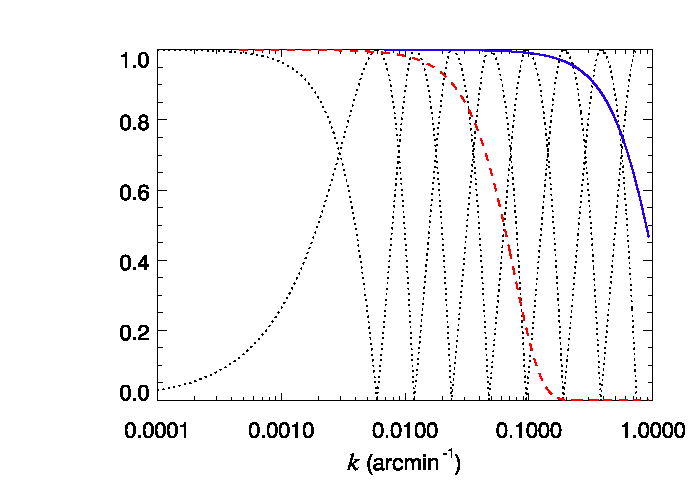}
  \end{center}
\caption{Illustration of needlet scales versus beams. In this plot we assume $N_1=5$ and $N_2=4$. The beam of \textsc{exp\oldstylenums{1}} channels (thick dashed red) and the beam of \textsc{exp\oldstylenums{2}} channels (thick solid blue) are overplotted on the needlet bands (thin dotted black). For each needlet band an ILC estimate is computed: in the first four bands the NILC processing can combine \textsc{exp\oldstylenums{1}} and \textsc{exp\oldstylenums{2}} channels whereas in the last four bands only \textsc{exp\oldstylenums{2}} channels are exploited by NILC.} 
\label{Fig:bandsvsbeams}
\end{figure}

\begin{table*}
\begin{flushleft}
\begin{tabular}{|p{2.4cm}|*{6}{l}}
\hline
{\bf Planck-like}   & 30 GHz & 90 GHz & 148 GHz & 219 GHz & 350 GHz & 545 GHz \\ 
\hline 
 $\theta$ [arcmin] & $32.65$ & $9.88$ & $7.18$ & $4.71$ & $4.5$ & $4.72$ \\
\hline
$\sigma_N \mathrm{(4~ surveys)}$ [$\mathrm{\mu K\cdot arcmin}$]  & $210$ & $57$ & $30$& $44$& $173$& $2485$\\
\hline
\end{tabular}
\begin{tabular}{|p{2.5cm}|*{3}{l}|}
\hline
{\bf ACT-like}            & 148 GHz & 219 GHz & 277 GHz \\
\hline 
$\theta$ [arcmin]  & $1.4$ & $1.0$ & $0.9$  \\
\hline
$\sigma_N$ [$\mathrm{\mu K\cdot arcmin}$]   & $30$ & $45$ & $60$ \\
\hline
\end{tabular}
\end{flushleft}
\caption{Specifications of both datasets constructed from simulations of \citet{2010ApJ...709..920S}. Beams and noise RMS for Planck-like maps refer to the values in \citet{planck2011-1.4}, \citet{planck2011-1.7} extrapolated to 4 full-sky surveys (28 months). ACT-like values refer to the 2008 ACT Southern survey \citep{2011ApJ...729...62D}.} \label{tab:table}
\end{table*}

\begin{figure*}
  \begin{center}
    \includegraphics[width=7cm]{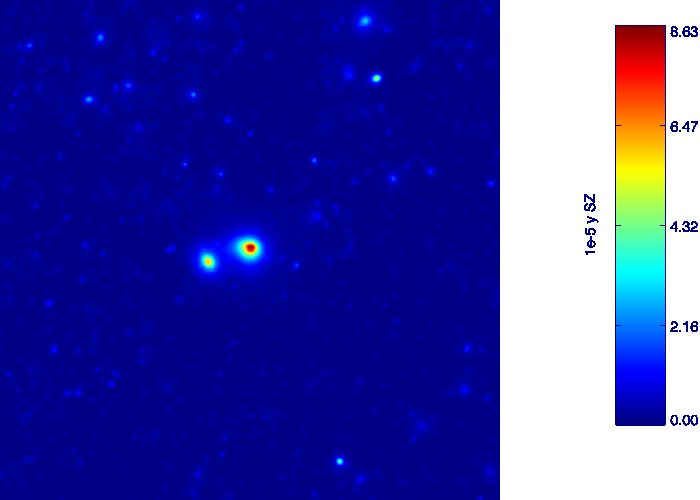}
    \includegraphics[width=7cm]{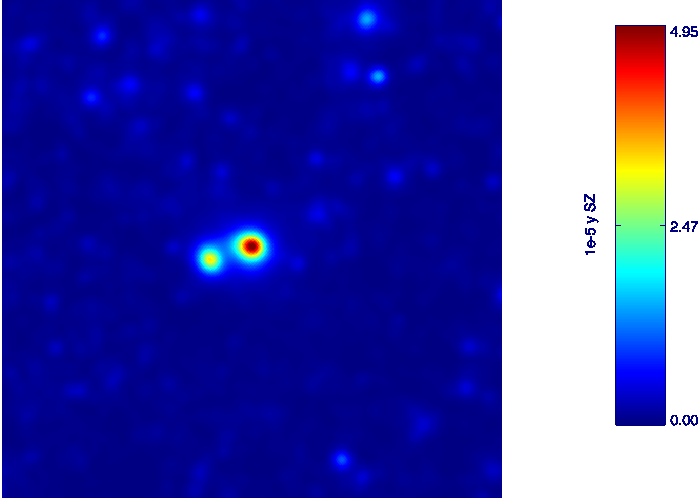}\\
    \includegraphics[width=7cm]{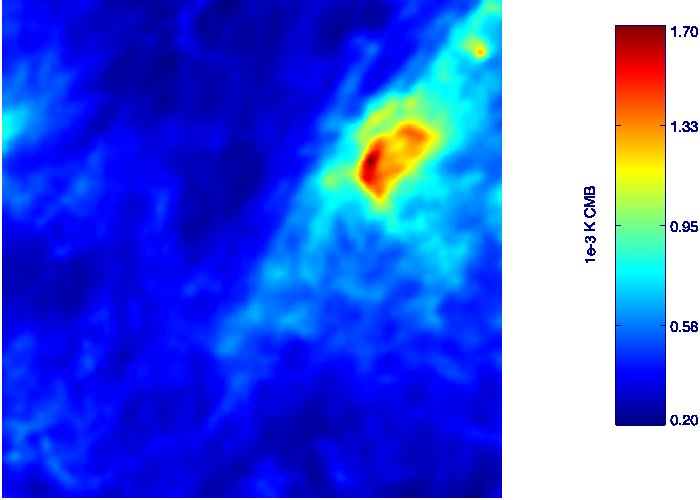}
    \includegraphics[width=7cm]{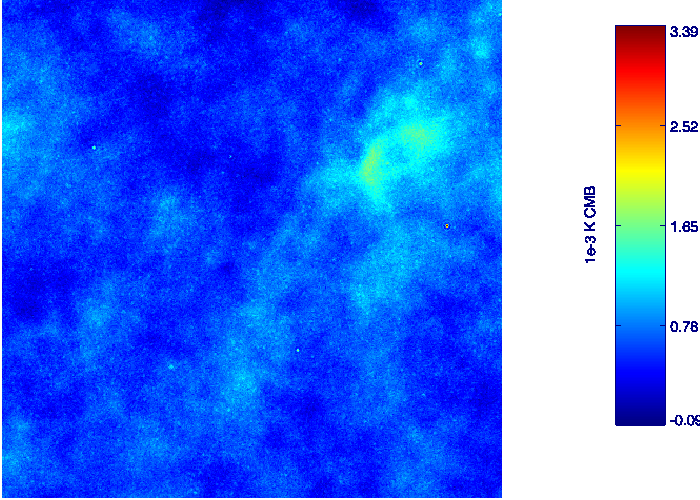}
  \end{center}
\caption{A $6.8^\circ\times 6.8^\circ$ patch of the simulated sky centered on the selected cluster 1 of Tab. \ref{tab:table_amas}. On the top panels: input thermal SZ smoothed to 3 arcmin resolution (left) and to 10 arcmin resolution (right). On the bottom panels: simulated Galactic dust at 277GHz (left) and ACT-like 277GHz channel-map (right).}
\label{Fig:inputtsz}
\end{figure*}

\begin{figure*}
  \begin{center}
    \includegraphics[width=5cm]{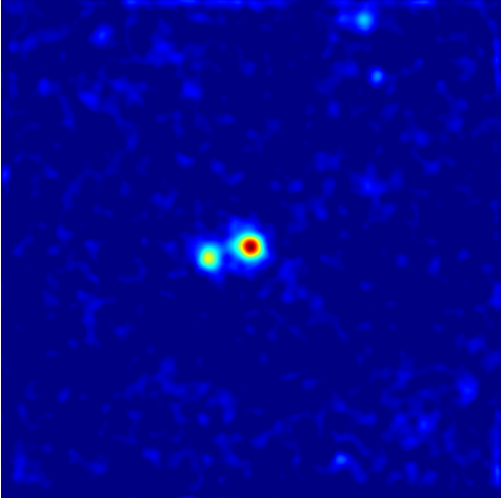}
    \includegraphics[width=5cm]{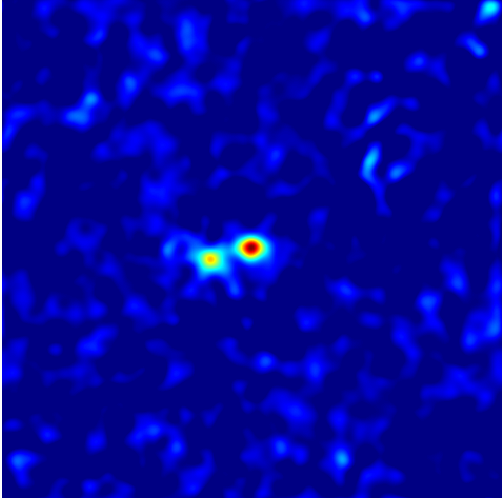}
    \includegraphics[width=6.98cm]{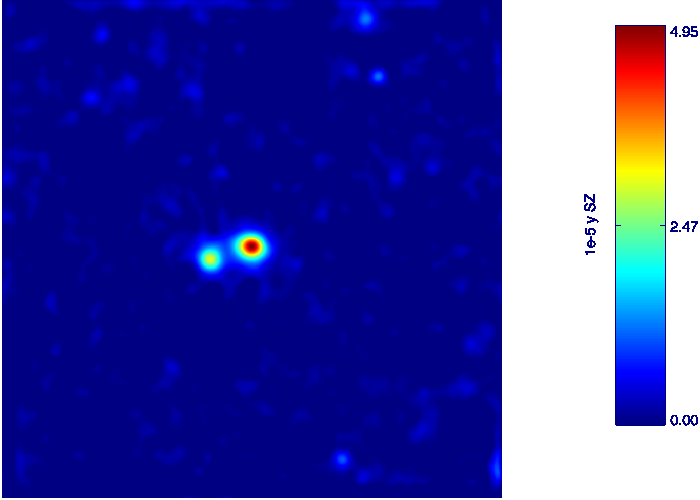}\\
   \includegraphics[width=5cm]{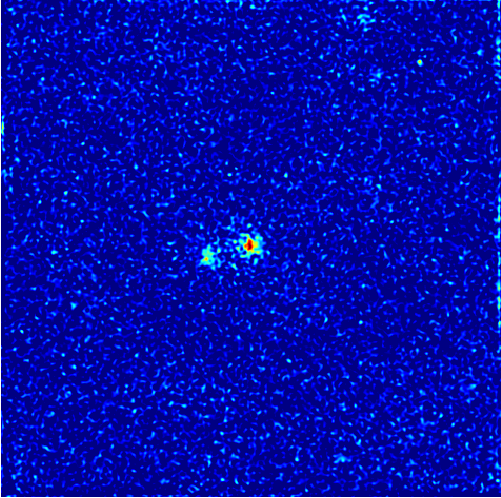}
    \includegraphics[width=5cm]{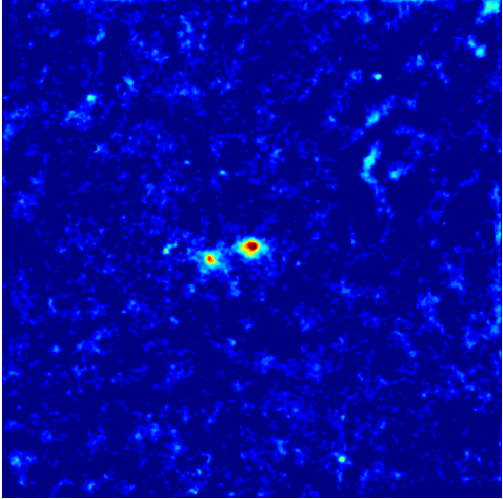}
   \includegraphics[width=6.98cm]{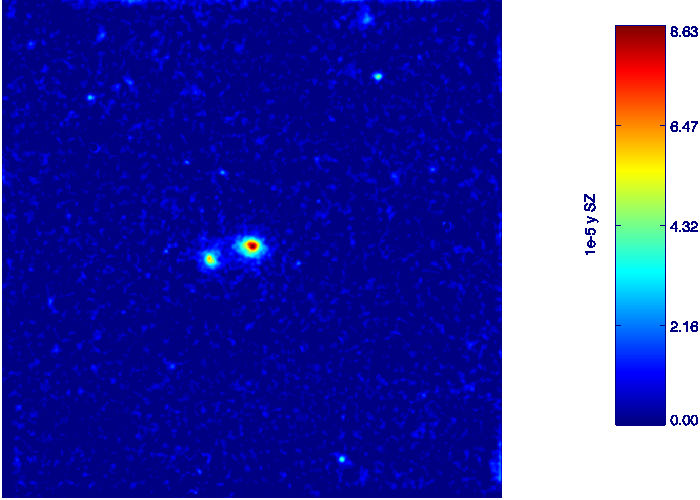}
  \end{center}
\caption{ NILC thermal SZ reconstruction of the selected cluster 1 of Tab. \ref{tab:table_amas} from different datasets at 10 arcmin (top panels) and 3 arcmin (bottom panels). Left panels: NILC TSZ reconstruction from Planck-like channels only. Middle panels: NILC TSZ reconstruction from ACT-like channels only. Right panels: NILC TSZ reconstruction from the combination of heterogeneous data, i.e. both Planck-like and ACT-like channels.} 
\label{Fig:channels}
\end{figure*}

\begin{figure*}
  \begin{center}
    \includegraphics[width=5cm]{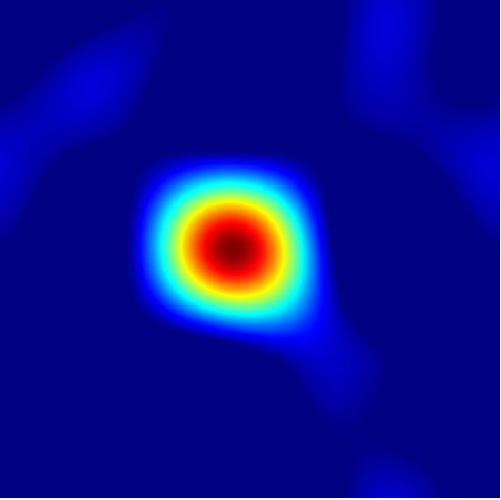}
    \includegraphics[width=5cm]{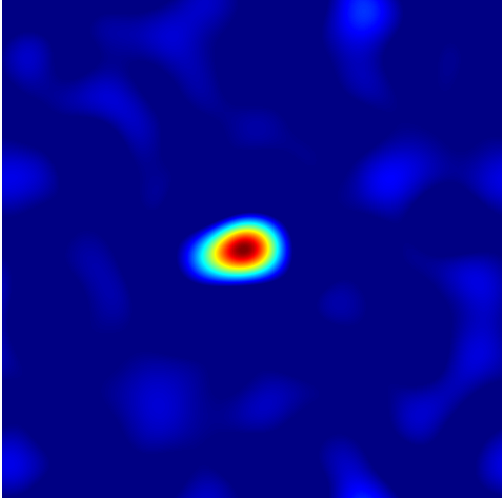}
    \includegraphics[width=5cm]{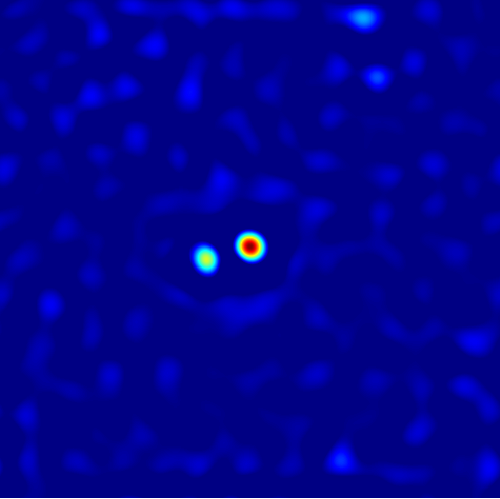}
    \includegraphics[width=5cm]{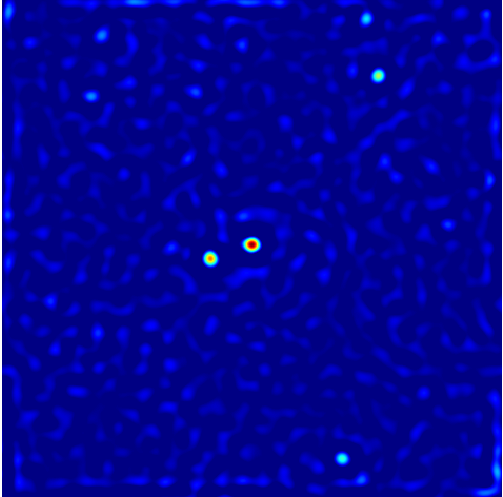}   
   \includegraphics[width=5cm]{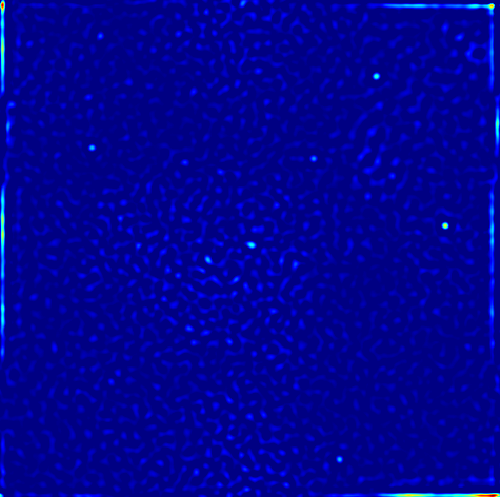}
   \includegraphics[width=5cm]{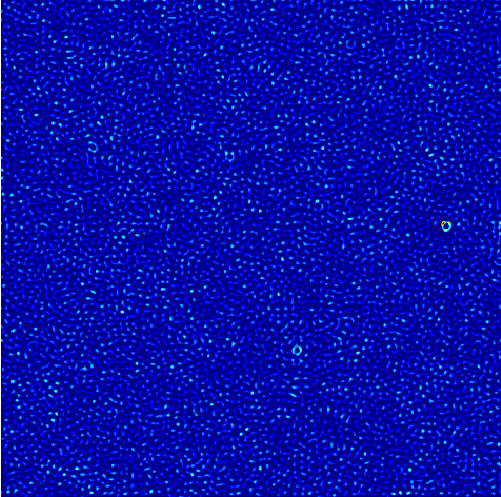}
    \includegraphics[width=5cm]{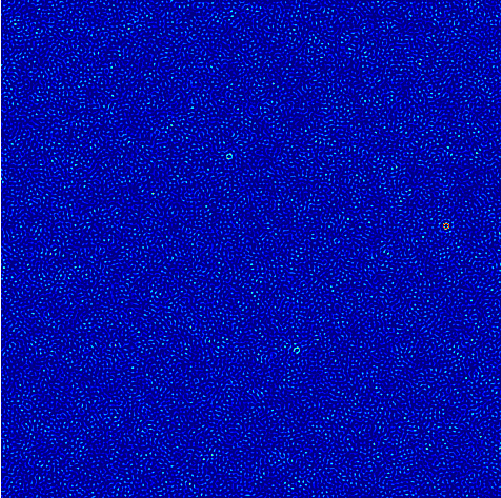}
    \includegraphics[width=6.5cm]{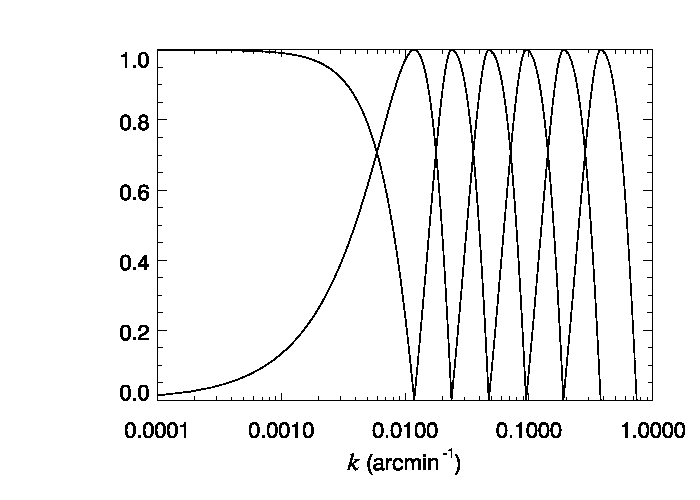}
  \end{center}
\caption{NILC reconstruction per scale (Planck-like/ACT-like combined) for the selected cluster $1$ of Tab. \ref{tab:table_amas}. Last panel: the associated needlet bands used for the reconstruction.} 
\label{Fig:perscale}
\end{figure*}

We may define $N_1+N_2$ needlet bands $h^{(j)}_k$ such that the beam of \textsc{exp\oldstylenums{1}}, $\theta_1$, can not probe the scales $k > k_{\max}^{(N_1)}$, where $k_{\max}^{(N_1)}$ is the highest Fourier mode probed by the $N_1^{\mathrm{th}}$ needlet band $h^{(N_1)}_k$ (Fig. \ref{Fig:bandsvsbeams}).

It is then possible to perform ILC filtering at each needlet scale $(j)$ by combining the channel-maps whose resolutions are compatible with the needlet band $h^{(j)}_k$ considered. Typically, in the bands $(j)$, where $1 \leq j \leq N_1$, the ILC TSZ output $\widehat{s}^{(j)}(p)$ is obtained from the combination of the $N_1+N_2$ maps of both \textsc{exp\oldstylenums{1}} and \textsc{exp\oldstylenums{2}}. Next, in the following bands $(j)$, where $N_1+1 \leq j \leq N_2$, the ILC combination is performed on the $N_2$ higher resolution maps of \textsc{exp\oldstylenums{2}} only, the maps of \textsc{exp\oldstylenums{1}} having no information at the scales covered by the $N_2$ latest bands to be taken into account by the ILC combination. Formally, for all $j \leq N_1$ the ILC estimate is computed from the combination of both the \textsc{exp\oldstylenums{1}} dataset and the \textsc{exp\oldstylenums{2}} dataset  
\ba
\widehat{s}^{(j)}(p) &=& w^{(j)}_1 x_{1 ~\mathrm{\textsc{exp\oldstylenums{1}}}}^{(j)}(p)+ ... + w^{(j)}_{N_1} x_{N_1 ~\mathrm{\textsc{exp\oldstylenums{1}}}}^{(j)}(p) \cr
& & + w^{(j)}_{N_1+1} x_{N_1+1 ~\mathrm{\textsc{exp\oldstylenums{2}}}}^{(j)}(p) + ... + w^{(j)}_{N_1+N_2} x_{N_1+N_2 ~\mathrm{\textsc{exp\oldstylenums{2}}}}^{(j)}(p),\nonumber
\ea
while for all $j > N_1$ the ILC is computed from the combination of the \textsc{exp\oldstylenums{2}} dataset only
\ba
\widehat{s}^{(j)}(p) &=& w^{(j)}_{N_1+1} x_{N_1+1 ~\mathrm{\textsc{exp\oldstylenums{2}}}}^{(j)}(p) + ... + w^{(j)}_{N_1+N_2} x_{N_1+N_2 ~\mathrm{\textsc{exp\oldstylenums{2}}}}^{(j)}(p).\nonumber
\ea

Therefore, NILC appears as a tunable multiresolution tool to aggregate multiple experiments with various specifications: larger angular scales are exploited to remove contaminants (Galactic foregrounds, etc) from the NILC combination of a large number of channels (both \textsc{exp\oldstylenums{1}} and \textsc{exp\oldstylenums{2}}), whereas smaller angular scales are exploited to reconstruct SZ effect at higher resolution from the combination of the high resolution \textsc{exp\oldstylenums{2}} channels only.


\section{Results}\label{sec:results}

\subsection{Sky simulations}

We now illustrate our method on the high resolution simulations of the microwave sky\footnote{The simulation can be found at http://lambda.gsfc.nasa.gov/toolbox/tb\_cmbsim\_ov.cfm} generated by \citet{2010ApJ...709..920S}. 
For this study the simulated maps have been degraded from their original HEALPix resolution nside=8192 to a HEALPix resolution nside=4096. The simulated all-sky emission is generated along six frequency channels ($30$ GHz, $90$ GHz, $148$ GHz, $219$ GHz, $277$ GHz and $350$ GHz). We have made use of the CMB emission, the thermal SZ effect, the Galactic dust emission, and the emissions from infrared and radio galaxies. All these components are coadded in each frequency channel to provide six sky maps. From these simulations we have created two datasets, a Planck-like set of maps and an ACT-like set of maps. The Planck-like and the ACT-like experimental characteristics are defined in Tab. \ref{tab:table}. Planck-like maps are constructed by smoothing the simulated sky maps, except the $277$ GHz sky map, to the beams given in  Tab. \ref{tab:table} and by adding a white noise map from the RMS levels of Tab. \ref{tab:table}. At this stage we have five Planck-like channel-maps but we also created a sixth map by extrapolating the $350$ GHz simulated components to $545$ GHz.  We have used for this extrapolation the temperature and spectral index of Galactic dust computed in \citet{planck2011-7.12}, and the temperature and spectral index of infrared sources computed in \citet{planck2011-6.6}. 
ACT-like maps are constructed by smoothing the $148$ GHz, $219$ GHz, and $277$ GHz simulated sky maps to the beams given in  Tab. \ref{tab:table} and by adding a white noise map at RMS values of Tab. \ref{tab:table}.
%
\begin{figure*}
  \begin{center}
    \includegraphics[width=8cm]{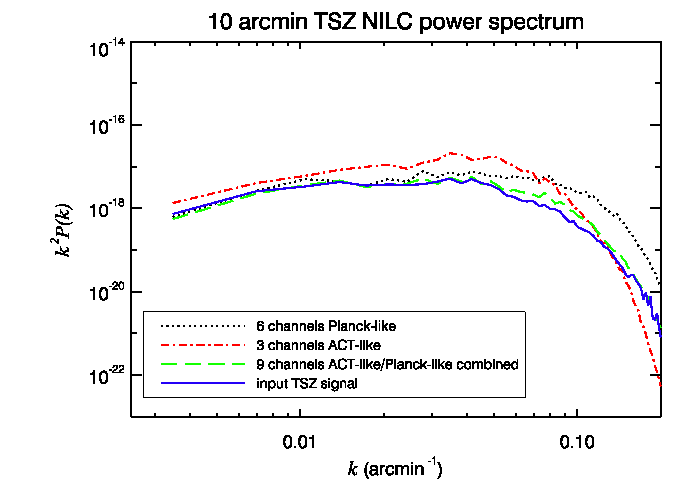}
    \includegraphics[width=8cm]{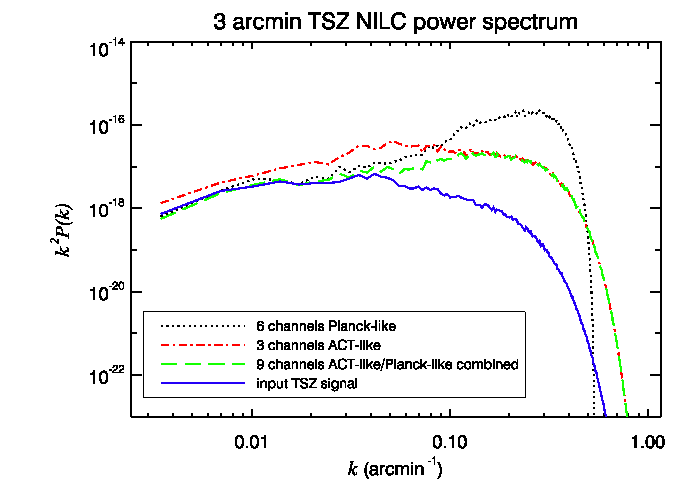}
    \includegraphics[width=8cm]{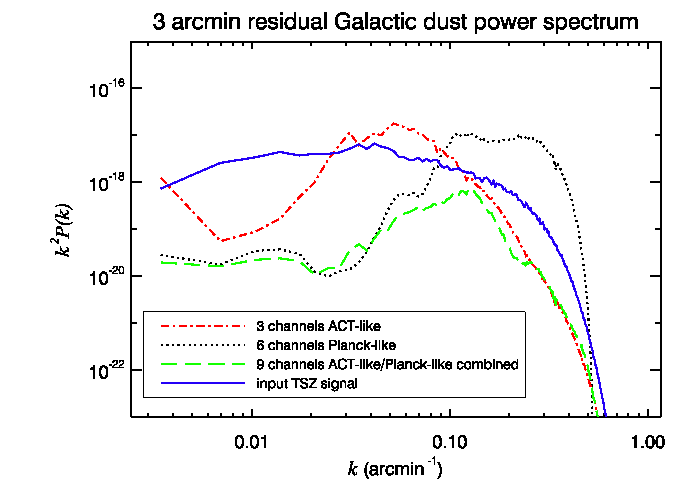}
    \includegraphics[width=8cm]{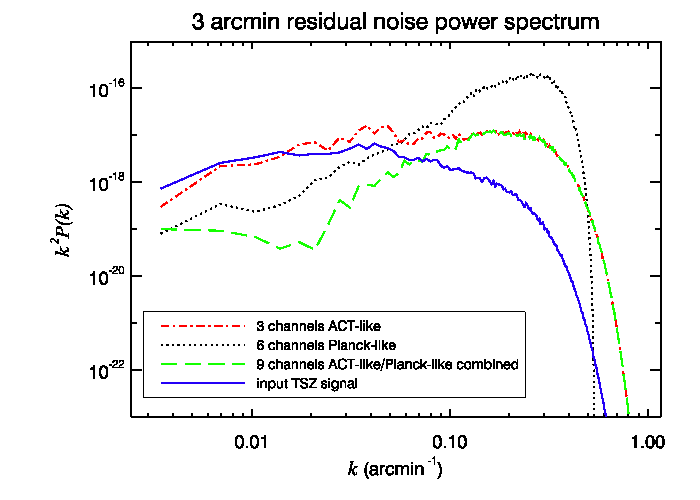}
  \end{center}
\caption{NILC thermal SZ power spectra from different datasets for $10$ arcmin reconstruction (top left panel) and $3$ arcmin reconstruction (top right panel) versus input power spectrum (selected cluster 1 of Tab. \ref{tab:table_amas}).  At the smallest scales, the Planck-like TSZ spectrum is noisy compared to the ACT-like TSZ power spectrum and limited to the instrumental beam. The joint Planck-like/ACT-like TSZ spectrum matches very well the input thermal SZ spectrum at $10$ arcmin resolution. 
Residual Galactic dust power spectra (bottom left panel) and residual instrumental noise power spectra (bottom right panel) for the $3$ arcmin TSZ reconstruction are also plotted for comparison. At the largest scales, the spectrum of the reconstructed TSZ from Planck-like data is less contaminated by the residual Galactic dust than the TSZ power spectrum from ACT-like data because of the higher number of channels combined by ILC. The joint Planck-like/ACT-like reconstruction shows an even better reduction of the residual dust power. 
 } 
\label{Fig:spectra}
\end{figure*}

\begin{table*}
\begin{flushleft}
\begin{tabular}{|p{2.5cm}|*{7}{c}|}
\hline
   &  z & R.A. & Dec. &$M_{200}$ & $R_{200}$ & $\theta_{200}$ &  $Y_{200}$\\
   &   & [degree] & [degree] &$[10^{14} M_\odot]$ & [Mpc] & [arcmin] &  [$\mathrm{arcmin}^2$]\\
\hline 
cluster 1 & $0.04$ & $39.6$ & $15.8$ & $7.3$& $1.82$ & $40.7$ & $0.0299$\\
\hline
cluster 2 &  $0.95$&  $2.5$ & $78.0$ & $9.7$& $1.45$& $3.0$ & $0.0007$\\
\hline
\end{tabular}
\end{flushleft}
\caption{Two selected clusters from the SZ Halo catalog of the simulation of \citet{2010ApJ...709..920S}. For each cluster, the table collects its redshift, its right ascension (R.A.) and its declination (Dec.), its mass, its radius, its angular size on the sky, and its integrated Compton value within $R_{200}$.}  
\label{tab:table_amas}
\end{table*}

In the next sections, we apply NILC on three sets of data (Planck-like dataset, ACT-like dataset, and joint Planck-like/ACT-like dataset) to retrieve the TSZ signal on $7^\circ\times 7^\circ$ patches of the sky centered on a chosen cluster. The TSZ reconstruction will be performed at $10$ and $5$ arcmin resolutions\footnote{The smallest resolution that we can reach when NILC is applied on the single Planck-like dataset is about $5$ arcmin (see Tab. \ref{tab:table}).} when  NILC is applied on the Planck-like dataset, whereas the reconstruction will be performed at $10$ and $3$ arcmin resolutions when NILC is applied either on the ACT-like dataset or on the joint Planck-like/ACT-like dataset. The reconstruction will be illustrated on two patches of the sky centered on two selected clusters given in Tab. \ref{tab:table_amas} (an extended one with $\theta_{200}=40.7$ arcmin and a compact one with $\theta_{200}=3$ arcmin) and then generalised to a sample of selected clusters. Here we define ${\theta_{200} = R_{200}/D_A}$ where $R_{200}$ is the radius in which the total density contrast of the cluster is ${\delta=200}$, as compared to the critical density of the Universe at the cluster redshift, and $D_A$ is the angular diameter distance to the cluster.

\subsection{NILC Thermal SZ reconstruction from combined ACT-like/Planck-like data}
 
\begin{figure*}
  \begin{center}
    \includegraphics[width=5.8cm]{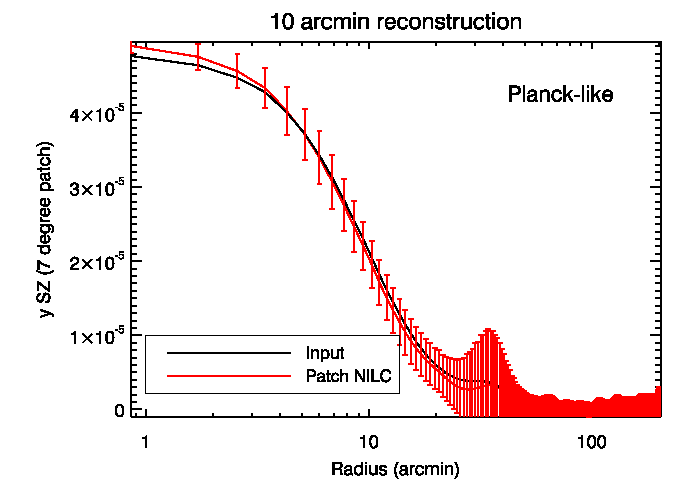}
    \includegraphics[width=5.8cm]{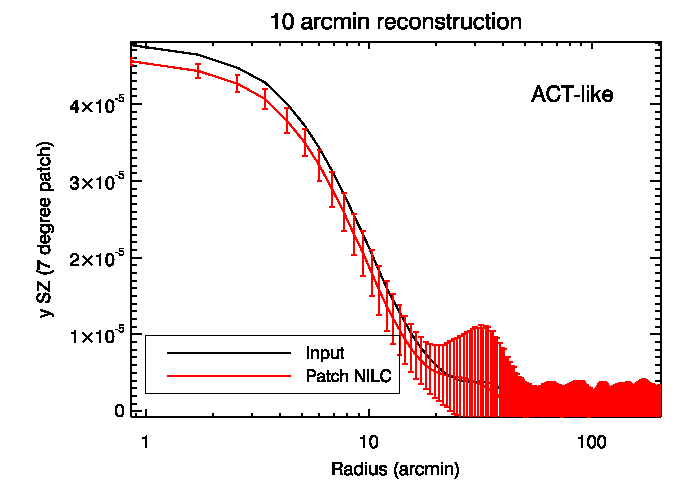}
    \includegraphics[width=5.8cm]{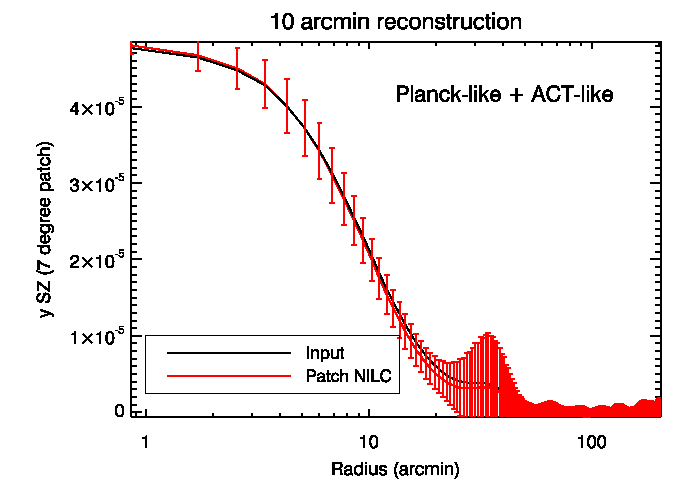}
    \includegraphics[width=5.8cm]{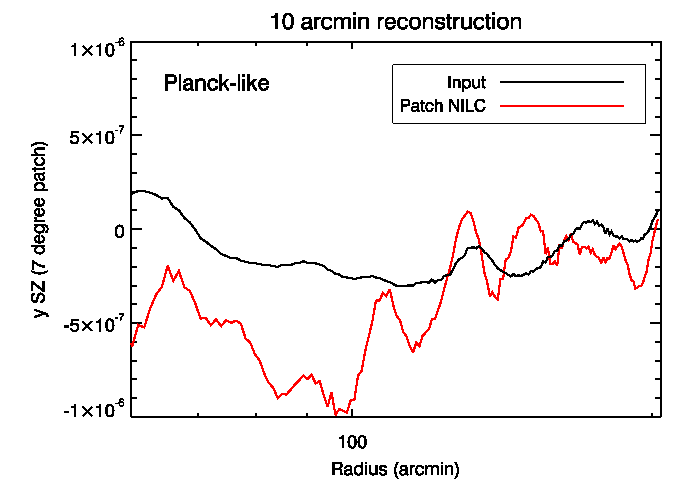}   
   \includegraphics[width=5.8cm]{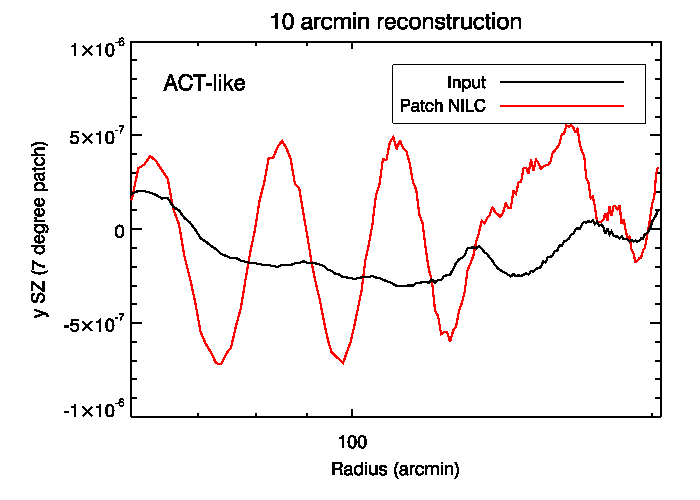}
   \includegraphics[width=5.8cm]{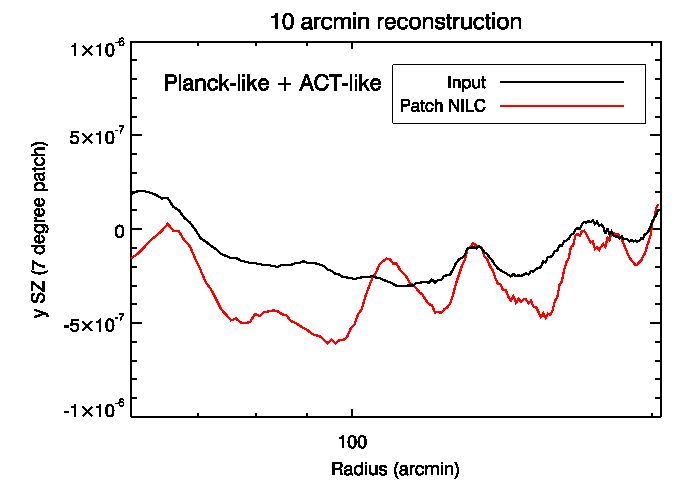}
  \end{center}
\caption{NILC TSZ profiles of the selected cluster $1$ of Tab. \ref{tab:table_amas} (10 arcmin TSZ reconstruction). Planck-like (left panels), ACT-like (middle panels), Planck-like/ACT-like combined (right panels).} 
\label{Fig:profils10}
\end{figure*}


\begin{figure*}
  \begin{center}
    \includegraphics[width=5.5cm]{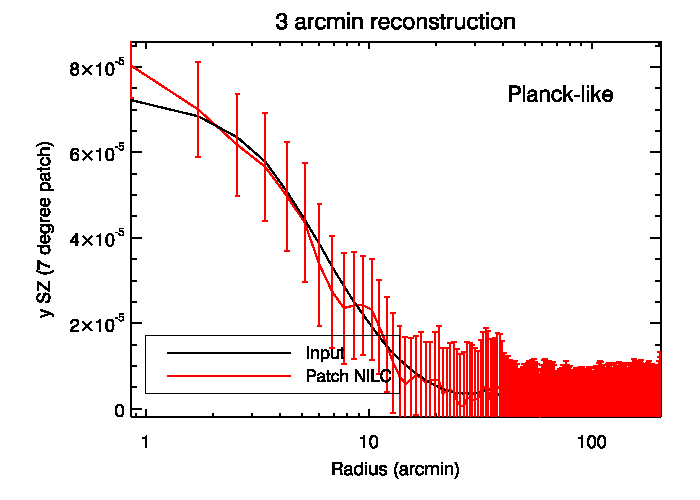}
    \includegraphics[width=5.5cm]{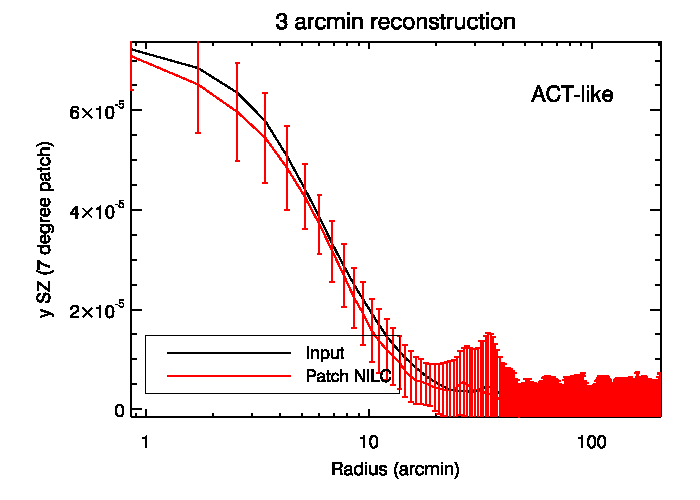}
    \includegraphics[width=5.5cm]{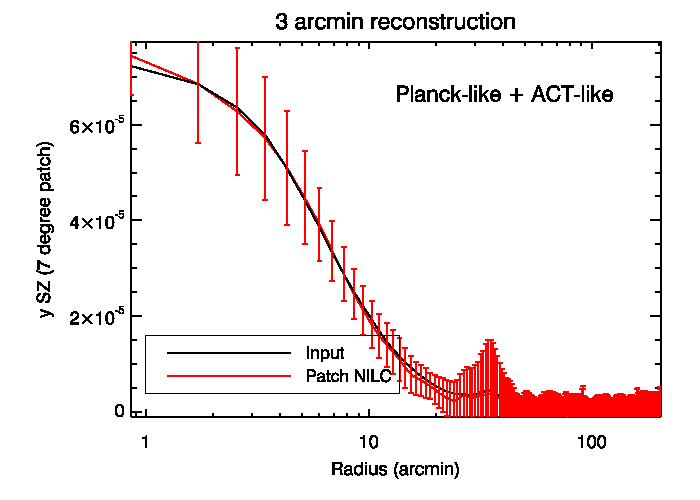}\\
   \includegraphics[width=5.5cm]{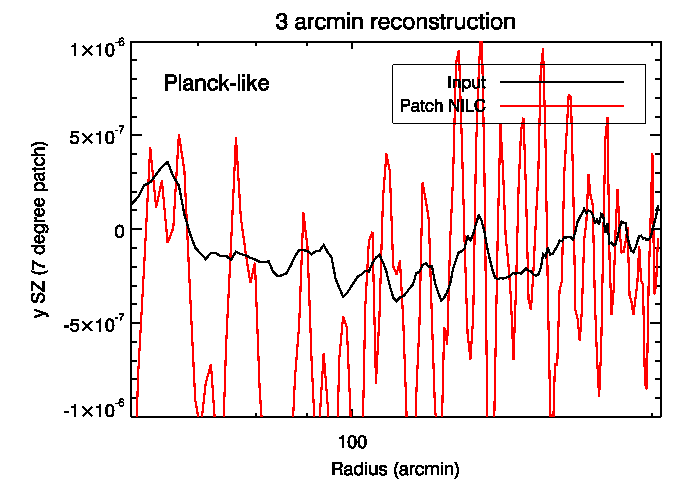}
   \includegraphics[width=5.5cm]{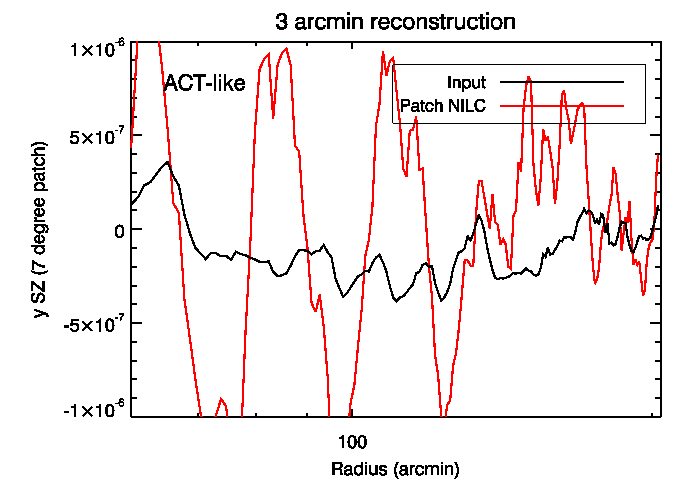}
   \includegraphics[width=5.5cm]{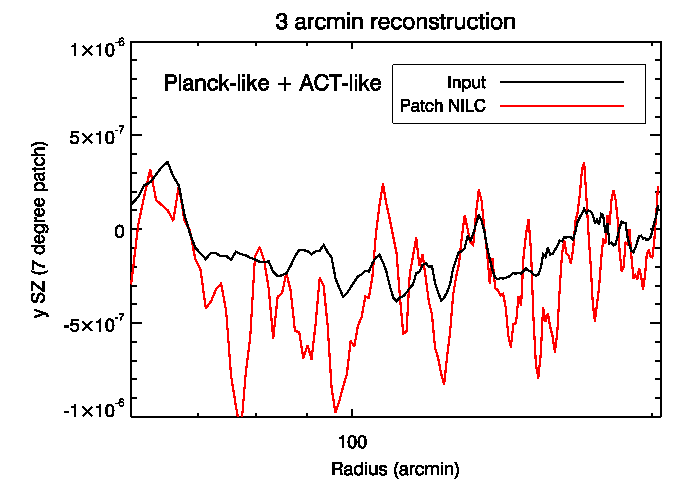}
  \end{center}
\caption{NILC TSZ profiles of the selected cluster $1$ of Tab. \ref{tab:table_amas} (3 arcmin TSZ reconstruction). Planck-like (left panels), ACT-like (middle panels), Planck-like/ACT-like combined (right panels).} 
\label{Fig:profils3}
\end{figure*}

\begin{figure*}
  \begin{center}
    \includegraphics[width=8cm]{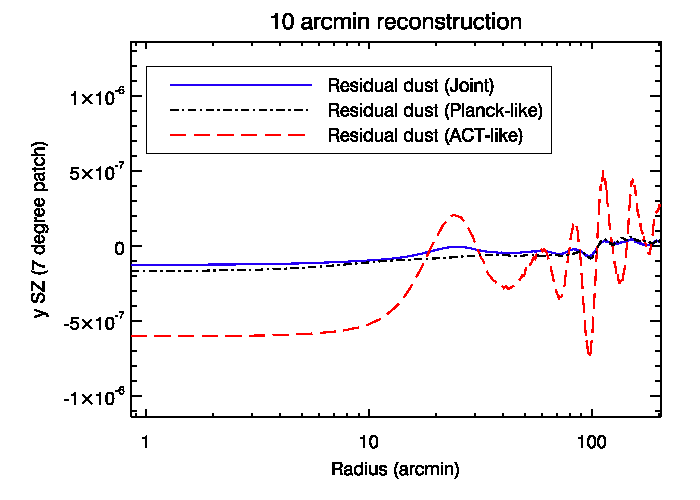}
    \includegraphics[width=8cm]{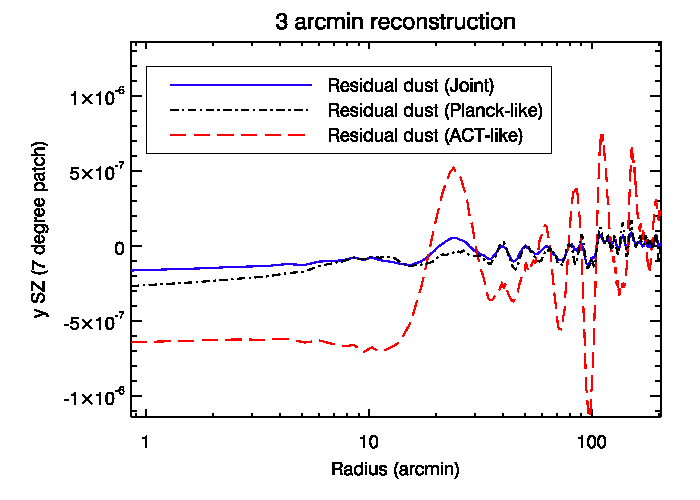}
    \includegraphics[width=8cm]{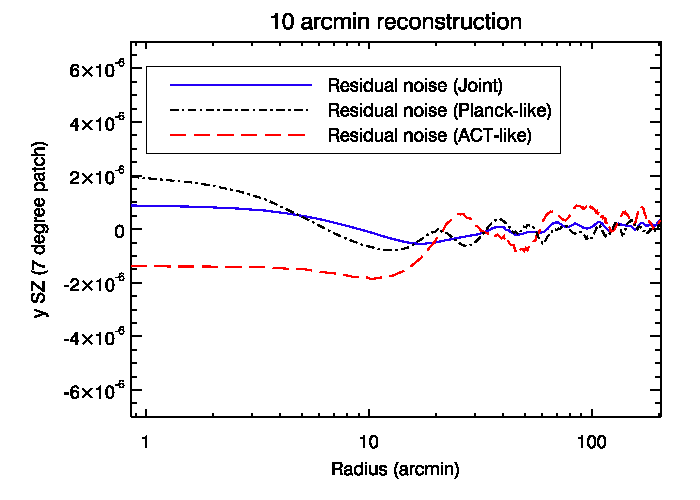}
    \includegraphics[width=8cm]{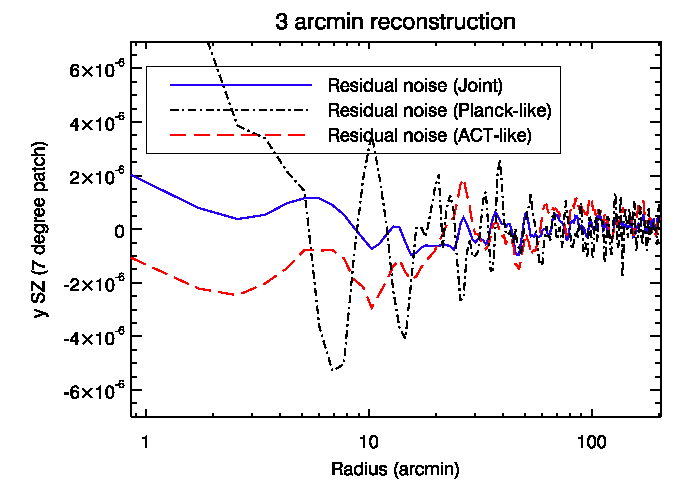}
  \end{center}
\caption{Residual dust profile (top panels) and residual noise profile (bottom panels) for the selected cluster $1$ of Tab. \ref{tab:table_amas}. Left panels: 10 arcmin TSZ reconstruction. Right panels: 3 arcmin TSZ reconstruction. Planck-like (dot-dashed black), ACT-like (long dashed red), Planck-like/ACT-like combined (solid blue).} 
\label{Fig:dustprofile}
\end{figure*}


The NILC reconstruction of the thermal SZ effect due to the selected extended cluster $1$ of Tab. \ref{tab:table_amas}, with typical angular size $\theta_{200}=40.7$ arcmin and located at a redshift $z=0.04$, is shown in the panels of Fig. \ref{Fig:channels}, whereas the input signal from the simulations is shown in Fig. \ref{Fig:inputtsz}.
We can see in the middle panels of Fig. \ref{Fig:channels} that the reconstructed TSZ map from ACT-like maps only is clearly more contaminated by foreground residuals (visible on the top right corner of the patches) than the one reconstructed from Planck-like maps only (left panels). Combining both datasets for the reconstruction further improves the removal of foreground residuals (right panels of Fig. \ref{Fig:channels}), by taking advantage of the large number of channels. Simultaneously, the combination of both Planck-like and ACT-like maps through Needlet ILC enables to reconstruct the thermal SZ signal at high resolution ($3$ arcmin in bottom right panel of Fig. \ref{Fig:channels}), which would be impossible from the use of Planck-like maps only since the beams are limited to $5$ arcmin. Note that on the bottom left panel the reconstructed TSZ map from Planck-like data actually has a resolution of $5$ arcmin. Moreover, Planck-like TSZ reconstruction at high resolution is noisy compared to ACT-like TSZ reconstruction because of the lower sensitivity. Once again, the joint exploitation of both datasets by NILC significantly reduces both the instrumental noise contamination and the sky residuals (right panels of Fig. \ref{Fig:channels}). These results clearly show that a reconstruction of thermal SZ from the combination of heterogeneous datasets (with different number of channels and different resolutions) benefits from the advantages of the TSZ reconstructions from each dataset independently while avoiding their drawbacks: low foreground contamination from Planck-like channels, high resolution and low noise contamination from ACT-like channels. For illustration, the NILC ``maps per scale'', which provide the reconstructed TSZ map by coaddition, can be seen in Fig. \ref{Fig:perscale}. Each patch corresponds to the ILC reconstruction at a given range of angular scales defined by the needlet spectral bands plotted in the last panel of Fig. \ref{Fig:perscale}. 

 The various power spectra of the reconstructed TSZ maps are shown in top panels of Fig. \ref{Fig:spectra}. The power spectrum of the NILC reconstructed TSZ map from the joint exploitation of both Planck-like maps and ACT-like maps best reconstructs the input TSZ power spectrum for the $10$ arcmin reconstruction (top left panel of Fig. \ref{Fig:spectra}) whereas the $3$ arcmin reconstruction becomes noisy at small scales $k > 0.05~ \mathrm{ arcmin}^{-1}$. As expected, on the top right panel of Fig. \ref{Fig:spectra}, NILC on Planck-like data better reconstructs the TSZ power spectrum at large scales ($k < 0.05~ \mathrm{ arcmin}^{-1}$) than ACT-like NILC because of the larger number of available channels used to remove the foreground contamination whereas NILC on ACT-like data performs better reconstruction at small scales than Planck-like NILC which lacks resolution and is noise-dominated. At large scales ($k < 0.05~ \mathrm{ arcmin}^{-1}$) NILC applied on joint datasets matches the Planck-like reconstruction whereas at small scales ($k > 0.05~ \mathrm{ arcmin}^{-1}$) it matches the ACT-like reconstruction, thus performing the best compromise. By applying the NILC weights onto the simulated frequency dust maps we are able to estimate the contamination level of Galactic dust emission in the reconstructed TSZ maps from each dataset. The power spectra of the residual Galactic dust emission for the $3$ arcmin reconstruction are plotted in the bottom left panel of Fig. \ref{Fig:spectra}: at scales $k < 0.1~ \mathrm{ arcmin}^{-1}$, the ACT-like TSZ reconstruction shows more residual Galactic dust than the Planck-like TSZ reconstruction, as it is expected from the reduced number of exploited channels. At smaller scales $k > 0.1~ \mathrm{ arcmin}^{-1}$, reconstruction in Planck-like case shows more Galactic dust residuals because the ILC weights are requested for minimizing the instrumental noise which becomes the dominant contaminant here, as it can be also seen in bottom right panel of Fig. \ref{Fig:spectra}. The TSZ NILC reconstruction from the joint exploitation of both ACT-like and Planck-like datasets reduces further the power of residual dust on all scales, matching the Planck-like reconstruction on large scales and the ACT-like reconstruction on small scales. In that case, on the largest scales, Galactic dust residuals are two orders of magnitude below the TSZ signal (bottom left panel of Fig. \ref{Fig:spectra}). The level of residual instrumental noise (bottom right panel of Fig. \ref{Fig:spectra}) is also reduced in all scales when reconstructing on joint ACT-like and Planck-like datasets, unlike single instrument reconstructions. However, instrumental noise still dominates the TSZ power spectrum on scales $k > 0.1~ \mathrm{ arcmin}^{-1}$.

In Fig. \ref{Fig:profils10} we plot the reconstructed TSZ cluster profiles versus the input TSZ cluster profile for the NILC reconstruction at $10$ arcmin. The analogue, at $3$ arcmin, is plotted in Fig. \ref{Fig:profils3}. We can see on the left panel the TSZ profile obtained from a reconstruction based on the Planck-like observations only, the middle panel shows the TSZ profile from  a reconstruction based on the ACT-like observations only, and the right panel exhibits the profile from a reconstruction jointly exploiting both ACT-like and Planck-like channels. We see how the combination of complementary datasets in the NILC filtering improves the TSZ profile reconstruction: less contamination is observed (right panels of Fig. \ref{Fig:profils10}) compared to single instrument reconstructions (left and middle panels). Moreover, the NILC filtering on joint datasets is able to recover the TSZ profile at high resolution (right panels of Fig. \ref{Fig:profils3}) where NILC on Planck-like data only fails (left panels of Fig. \ref{Fig:profils3}). 
We quantitatively measure the performance of the reconstruction of the SZ profile by the trade-off between the bias and the variance as a criterion. This trade-off is measured by the mean squared error (MSE)
 \ba\mathrm{MSE}~ = \sum_{\mathrm{ring}~ i}(\widehat{Y}_i-Y_i^{\mathrm{(in)}})^2+\sigma_i^2,\label{qqq:mse}\ea 
where $\widehat{Y}_i$ is the reconstructed profile, $Y_i^{\mathrm{(in)}}$ the input profile, and $\sigma_i$ the standard deviation (error bar) on the reconstructed profile for a ring $i$ of pixels. The standard deviation is given by \ba\label{qqq:stddev}\sigma_i = \sqrt{{1\over N_{\mathrm{p}}-1}\sum_{p~ \in~ \mathrm{ring}} \left(\widehat{y}(p) - \widehat{Y}_i\right)^2}\ea where $\widehat{y}$ is the reconstructed SZ map and $N_p$ the number of pixels in the ring $i$ considered. The results are listed in Tab. \ref{tab:chi2} for each reconstruction. 
\begin{table}
\begin{flushleft}
\begin{tabular}{|p{2.5cm}|*{4}{c}|}
\hline
 MSE ($10$ arcmin)  &  Planck-like & ACT-like & Joint \\
\hline 
  & $1.77\mathrm{e}$--$9$ & $3.82\mathrm{e}$--$9$ & $1.20\mathrm{e}$--$9$ \\
\hline
\hline
 MSE ($3$ arcmin)  &  Planck-like & ACT-like & Joint \\
\hline
& $2.70\mathrm{e}$--$8$ &  $9.58\mathrm{e}$--$9$ & $5.11\mathrm{e}$--$9$\\
\hline
\end{tabular}
\end{flushleft}
\caption{Performance of the profile reconstruction measured by the criterion Eq. (\ref{qqq:mse}). Whereas Planck-like reconstruction is better than ACT-like reconstruction at $10$ arcmin, it is the opposite for the $3$ arcmin reconstruction. For both output resolutions, the best reconstruction is performed from the joint Planck-like/ACT-like reconstruction.}  
\label{tab:chi2}
\end{table}
We see from this criterion that Planck-like $10$ arcmin reconstruction of the TSZ profile is better than ACT-like reconstruction whereas ACT-like NILC better reconstructs high resolution profile at $3$ arcmin than Planck-like NILC. For both output resolutions, the best reconstruction of the TSZ profile is performed by the joint Planck-like/ACT-like NILC. Table \ref{tab:chi2} indicates that Planck-like channels are exploited by NILC to remove sky contamination at large scales and lower resolution whereas high resolution ACT-like channels are exploited for reconstructing the SZ signal at higher resolution.  
The oscillations in the reconstructed profile (Figs. \ref{Fig:profils10} and \ref{Fig:profils3}) far from the center of the cluster (at a radius larger than $50$ arcmin) are due to residual temperature fluctuations of sky contaminants and instrumental noise residuals. We can see on the bottom panels of Fig. \ref{Fig:profils3} that the amplitude of oscillations is smaller when using both datasets for the reconstruction. Moreover, the typical wavelength of the oscillations is smaller in the Planck-like case than in the ACT-like case, indicating that Planck-like reconstruction is mostly contaminated by instrumental noise (small wavelength fluctuations) whereas ACT-like reconstruction is mostly contaminated by Galactic residuals (longer wavelength fluctuations). This is also confirmed by the right panels of  Fig. \ref{Fig:dustprofile} showing both residual Galactic dust in the profile reconstruction and residual noise with smaller wavelength fluctuations. These results illustrate how NILC applied jointly on Planck-like/ACT-like datasets is able to reconstruct the TSZ effect with a low sky residual contamination on the one hand (Fig. \ref{Fig:profils10}) and at a resolution beyond the beam of the Planck-like instrument on the other hand (Fig. \ref{Fig:profils3}).    

We also compute the contamination of Galactic dust and instrumental noise in the reconstructed TSZ profile by applying the NILC weights either onto the input dust maps of the simulation or onto the simulated noise maps. Top panels of Fig. \ref{Fig:dustprofile} represent the contamination of Galactic dust in the reconstructed TSZ profile for the three sets of data (Planck-like only, ACT-like only, and joint datasets). Again we see that Galactic dust contaminates more the ACT-like TSZ profile reconstruction with larger amplitude fluctuations than the Planck-like one because of the reduced number of exploited channels in the ACT-like case. The contamination of Galactic dust is clearly reduced further when performing NILC filtering jointly on both datasets. The residual instrumental noise in the TSZ profile on the bottom right panel of Fig. \ref{Fig:dustprofile} shows oscillations with larger amplitude in the Planck-like case than in the ACT-like case for the $3$ arcmin TSZ reconstruction, and the amplitude decreases when applying NILC jointly on both datasets.      

\begin{figure*}
  \begin{center}
    \includegraphics[width=5cm]{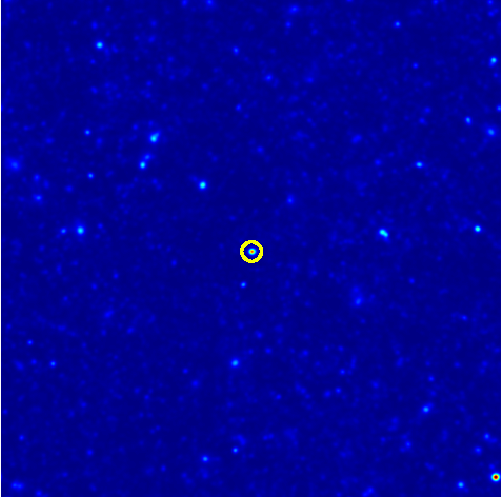}
    \includegraphics[width=6.98cm]{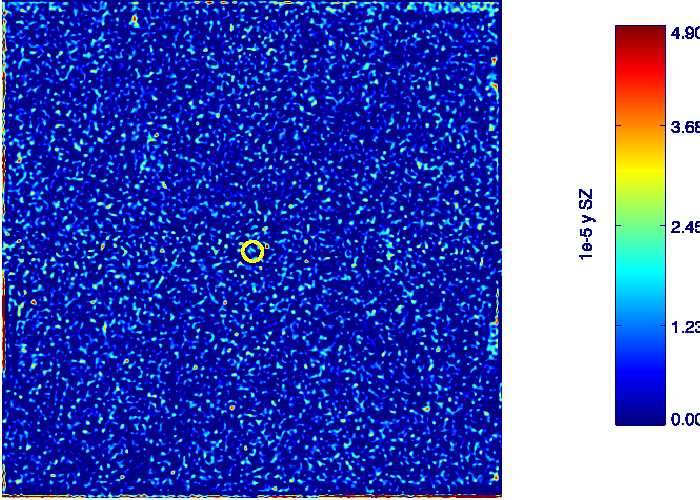}\\
    \includegraphics[width=5cm]{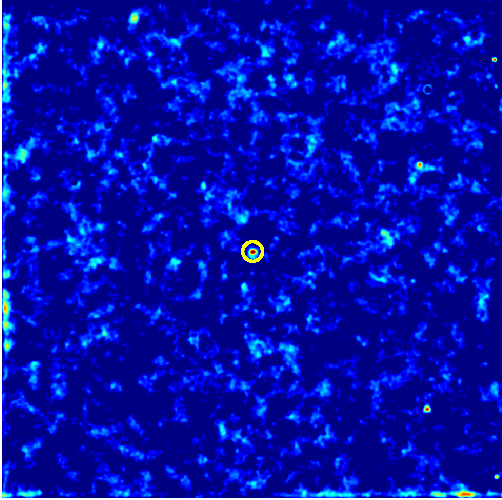}
    \includegraphics[width=6.98cm]{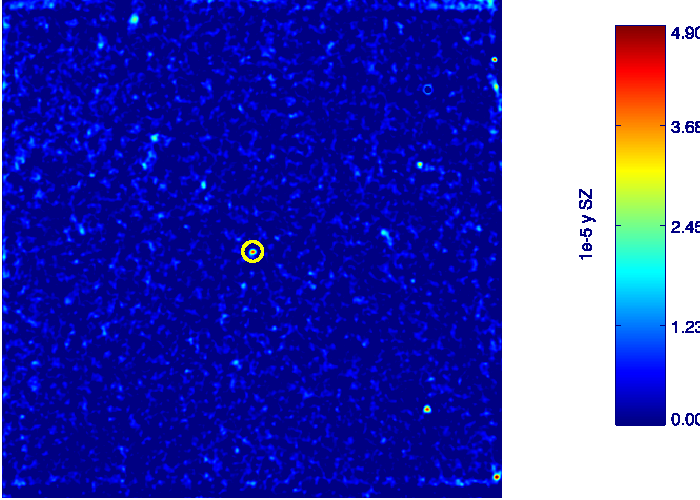}
  \end{center}
\caption{A $6.8^\circ\times 6.8^\circ$ patch of the simulated sky centered on the selected compact cluster 2 of Tab. \ref{tab:table_amas}. A yellow circle has been overplotted to indicate the location of cluster. Top left panel: Input thermal SZ of the compact cluster smoothed to 3 arcmin resolution. Top right panel: NILC TSZ reconstruction from Planck-like channels only (limited to 5 arcmin reconstruction), no TSZ detection, instrumental noise contaminated. Bottom left panel: NILC TSZ reconstruction from ACT-like channels only (3 arcmin reconstruction), TSZ detection, sky contaminated. Bottom right panel: NILC TSZ reconstruction from the combination of both Planck-like and ACT-like channels (3 arcmin reconstruction), TSZ detection with low contamination.} 
\label{Fig:smallcluster}
\end{figure*}

\subsection{Thermal SZ reconstruction of a compact cluster}

We now repeat the analysis on a more compact cluster. The compact cluster $2$ of Tab. \ref{tab:table_amas} is located at a redshift ${z\sim 1}$ with a typical angular size on the sky ${\theta_{200}= 3~ \mathrm{arcmin}}$. 

The reconstruction of the selected compact cluster is shown in Fig. \ref{Fig:smallcluster}. Whereas an ILC applied on Planck-like channels only is unable to reconstruct the TSZ signal from such a cluster (top right panel of Fig. \ref{Fig:smallcluster}), the same component separation applied on high-resolution ACT-like channel-maps successfully reconstructs the TSZ signal (bottom left panel of Fig. \ref{Fig:smallcluster}). However, the background residual noise from Galactic foregrounds is not successfully filtered. On the bottom right panel of Fig. \ref{Fig:smallcluster}, the wavelet ILC exploiting both Planck-like and ACT-like channel-maps is able to simultaneously reconstruct the TSZ signal from such a compact cluster and to successfully clean the background residual noise. 
 The joint exploitation of both datasets by a needlet ILC method shares with the ACT-like NILC the ability of reconstructing the TSZ effect from compact clusters of typical angular size below the beam of the single Planck-like instrument on the one hand, and shares with Planck-like NILC the ability of minimizing the sky residual contamination on the other hand.


\begin{figure*}
  \begin{center}
    \includegraphics[width=5.5cm]{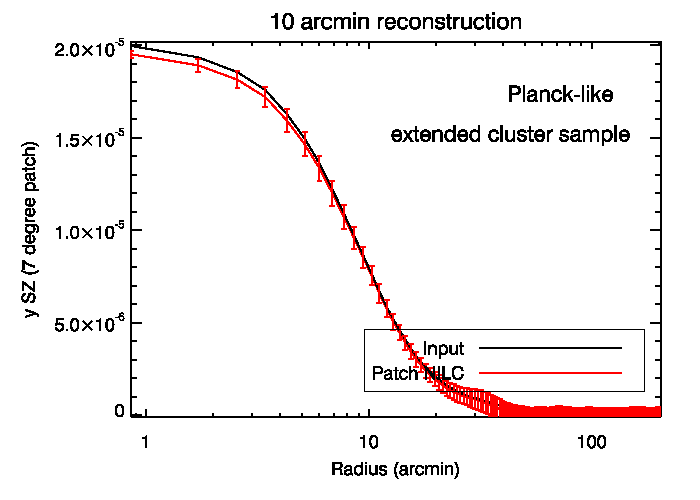}~
    \includegraphics[width=5.5cm]{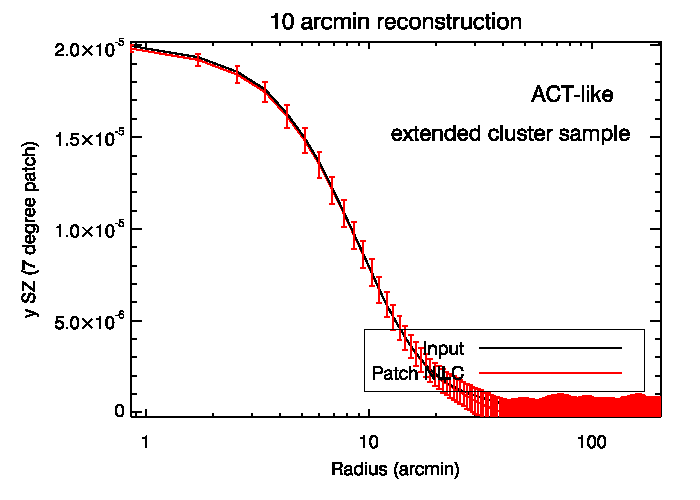}~
    \includegraphics[width=5.5cm]{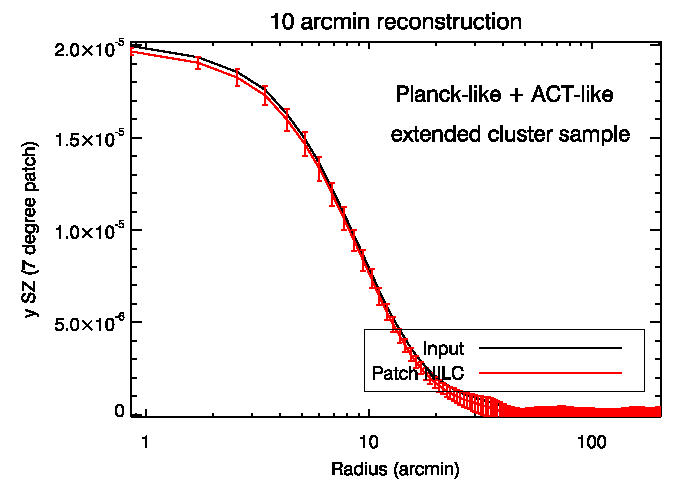}\\
     \includegraphics[width=5.5cm]{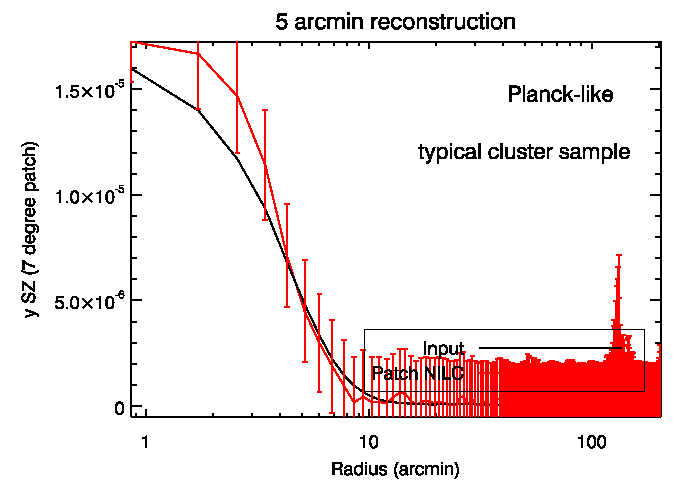}
    \includegraphics[width=5.5cm]{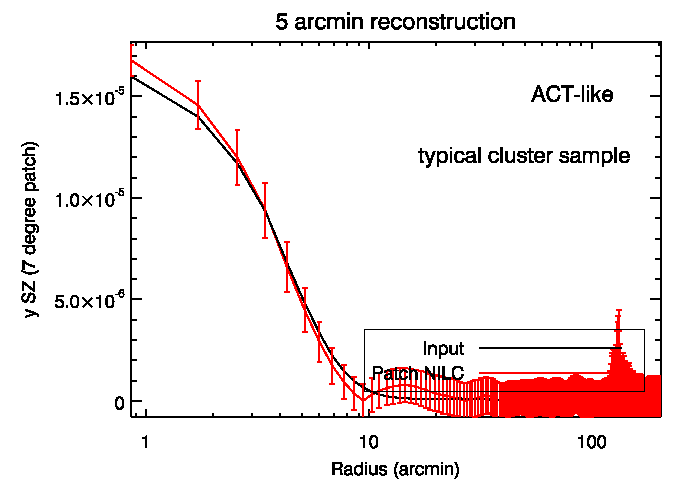}
    \includegraphics[width=5.5cm]{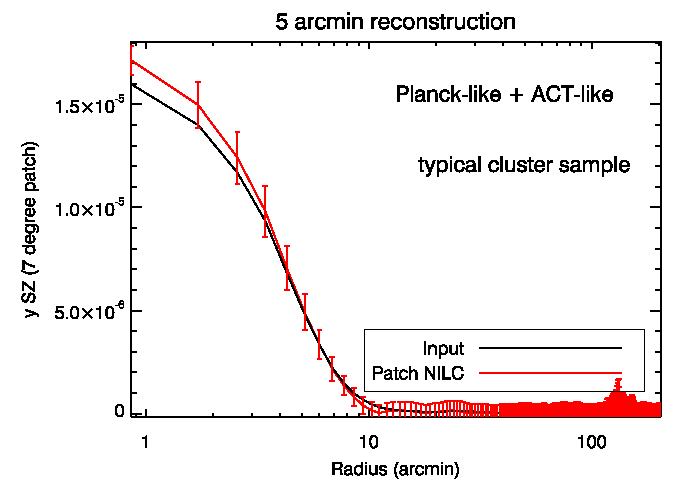}\\
     \includegraphics[width=5.5cm]{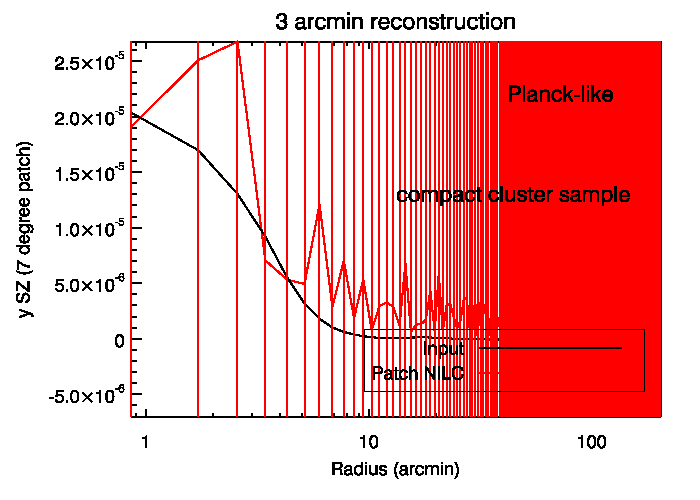}
    \includegraphics[width=5.5cm]{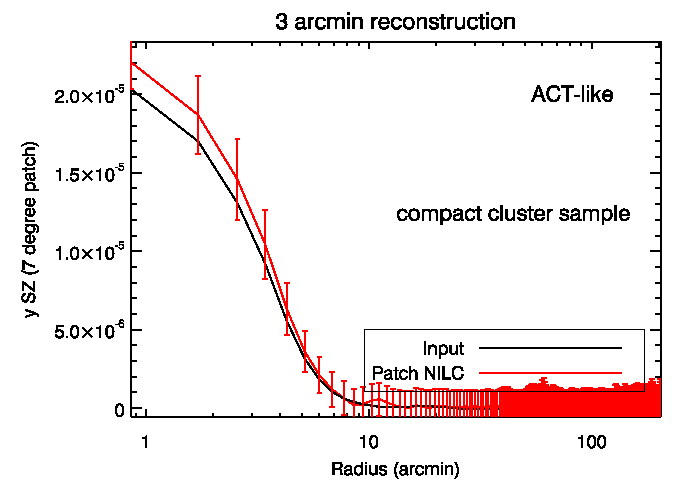}
    \includegraphics[width=5.5cm]{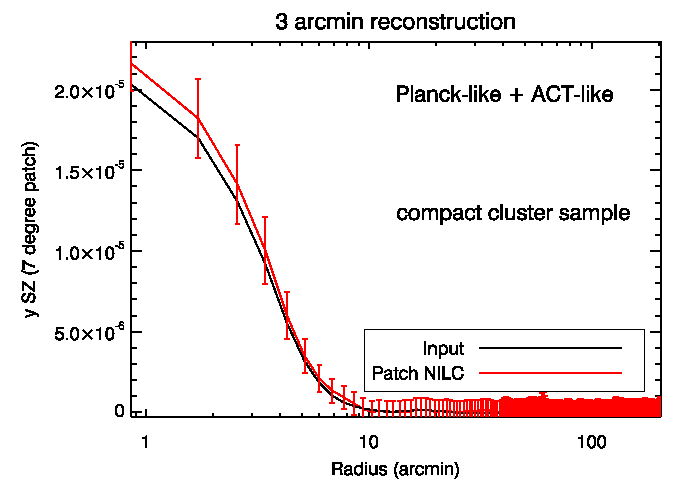}\\
  \end{center}
\caption{Cluster mean y-profiles for $7^\circ\times 7^\circ$ patches of the sky.  \underline{From left to right}: reconstructed using Planck-like channels, reconstructed using ACT-like channels, reconstructed using jointly Planck-like/ACT-like channels. \underline{From top to bottom}: sample of 30 extended clusters with ${15~ \mathrm{arcmin} < \theta_{200} < 40~ \mathrm{arcmin}}$, sample of 30 ``typical'' clusters with ${\theta_{200} \sim 5~ \mathrm{arcmin}}$, sample of 30 compact clusters with ${\theta_{200} \sim 3~ \mathrm{arcmin}}$.} 
\label{Fig:sampleflux}
\end{figure*}

\begin{figure*}
  \begin{center}
    \includegraphics[width=5.5cm]{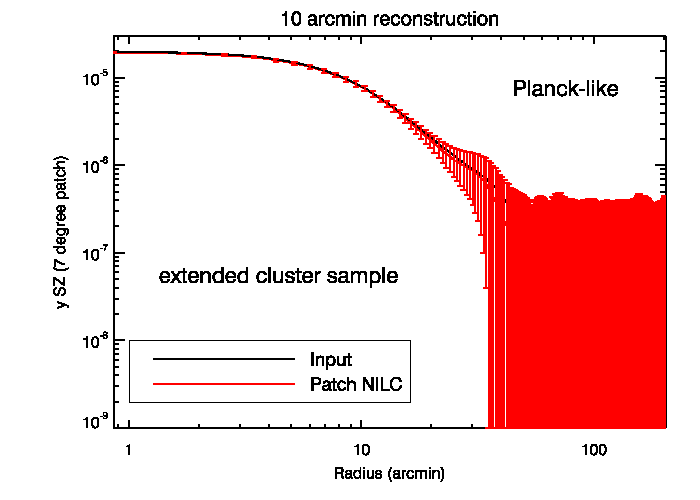}~
    \includegraphics[width=5.5cm]{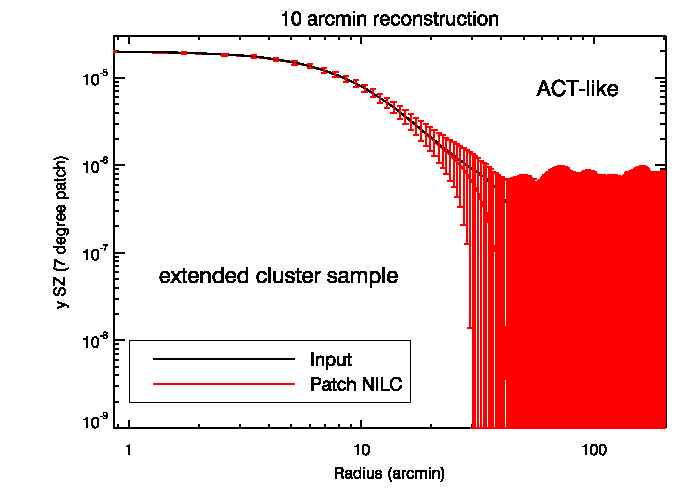}~
    \includegraphics[width=5.5cm]{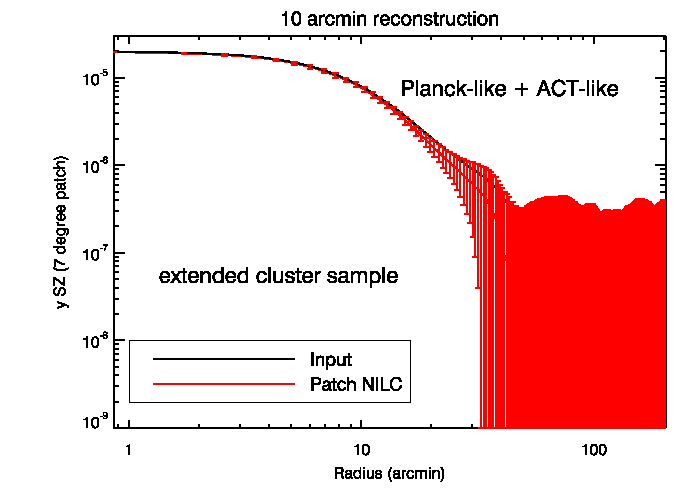}\\
     \includegraphics[width=5.5cm]{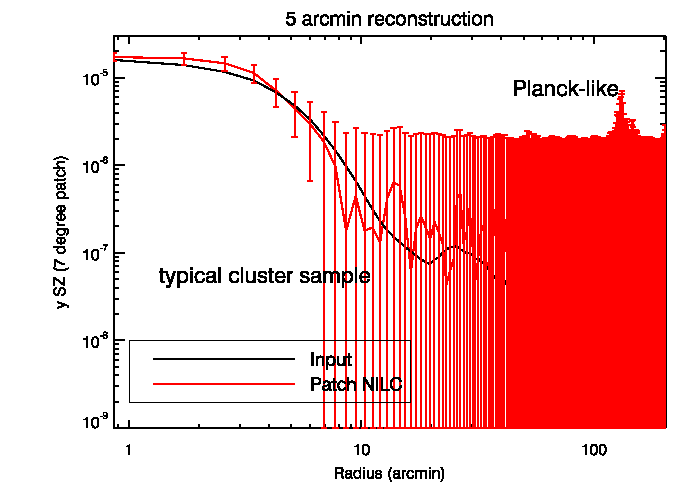}
    \includegraphics[width=5.5cm]{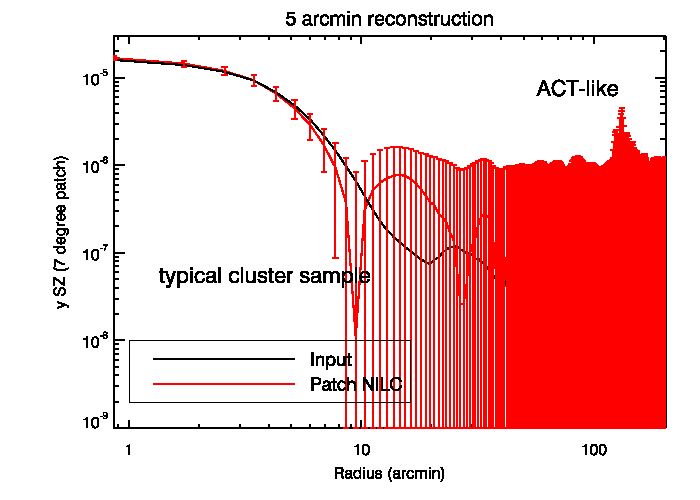}
    \includegraphics[width=5.5cm]{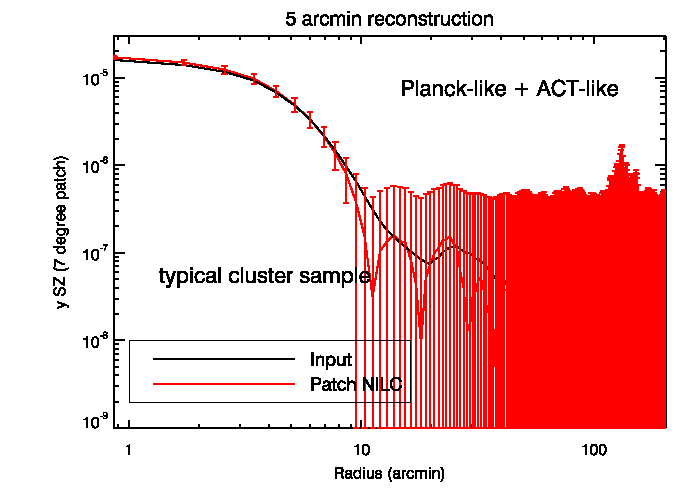}\\
     \includegraphics[width=5.5cm]{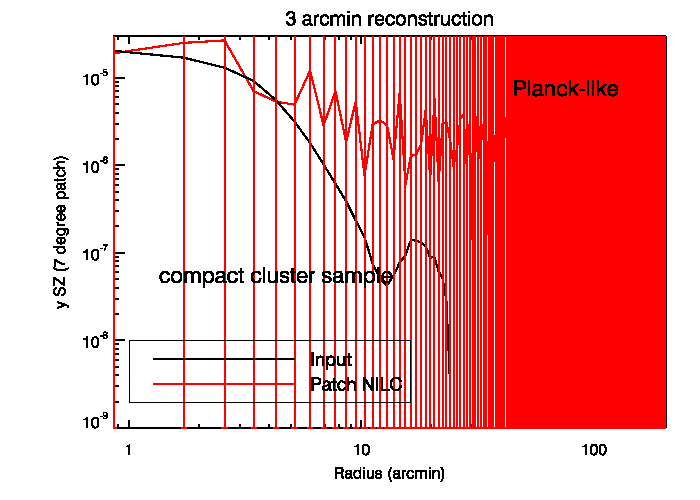}
    \includegraphics[width=5.5cm]{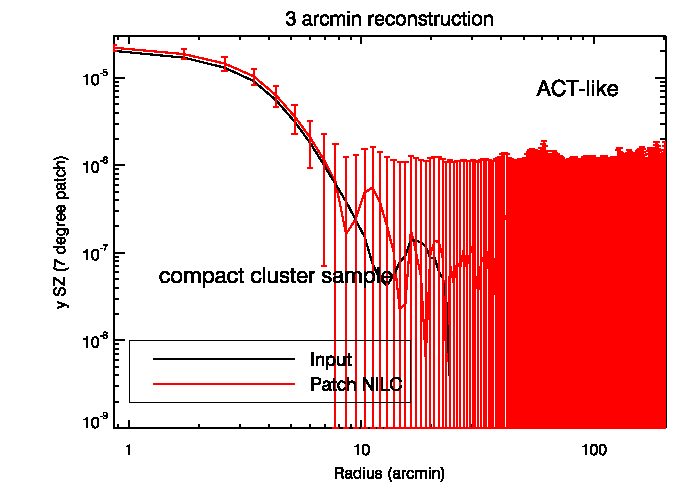}
    \includegraphics[width=5.5cm]{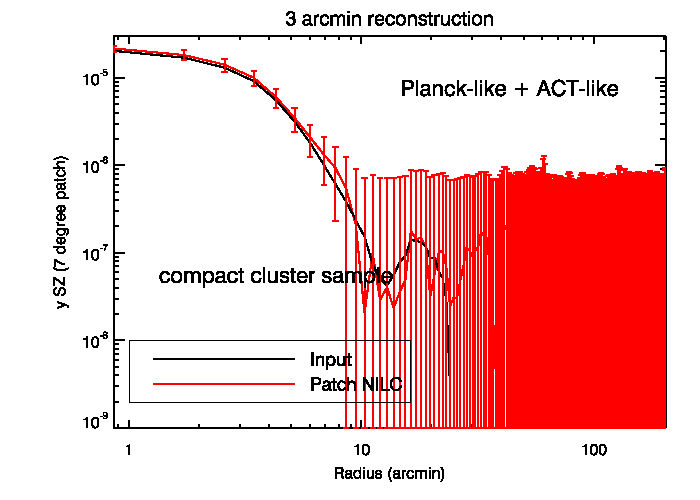}\\
  \end{center}
\caption{Same as Fig. \ref{Fig:sampleflux} in logarithmic scale. } 
\label{Fig:samplefluxlog}
\end{figure*}

\subsection{Statistical analysis on a sample of SZ galaxy clusters}

We now perform the same analysis on a selected sample of $90$ SZ clusters from the catalog of \citet{2010ApJ...709..920S} consisting in three sub-samples of $30$ clusters each: a sub-sample of $30$ extended clusters with ${15~ \mathrm{arcmin} < \theta_{200} < 40~ \mathrm{arcmin}}$, a sub-sample of $30$ ``typical'' clusters with ${\theta_{200} \sim 5~ \mathrm{arcmin}}$, and a sub-sample of $30$ compact clusters with ${\theta_{200} \sim 3~ \mathrm{arcmin}}$. Their different locations on the sky allow us to explore different levels of sky background. The mean TSZ profiles are shown in {Fig. \ref{Fig:sampleflux}}. 

\begin{figure*}
  \begin{center}
    \includegraphics[width=10cm]{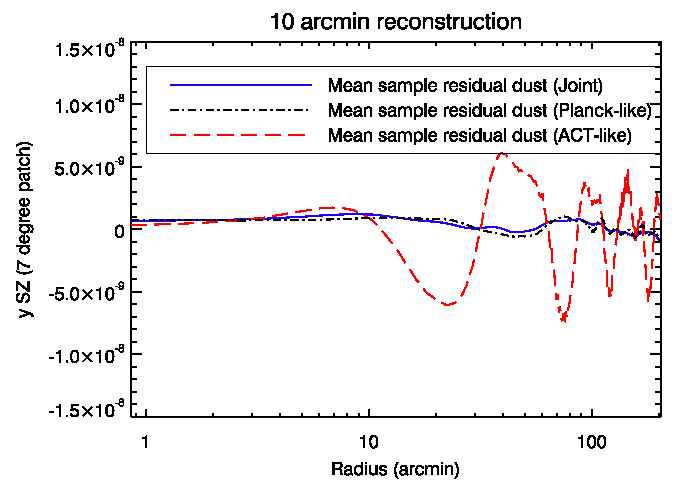}
      \end{center}
\caption{Residual dust mean profile (extended cluster sample). Planck-like (dot-dashed black), ACT-like (long dashed red), Planck-like/ACT-like combined (solid blue).} 
\label{Fig:sampledustflux}
\end{figure*}

On the sub-sample of extended clusters (top panels of Fig. \ref{Fig:sampleflux}) the reconstruction of the TSZ profile by NILC is reliable for the three sets of data (Planck-like only, ACT-like only, and joint datasets) with the ACT-like reconstruction slightly more contaminated by sky residual fluctuations at large radii. This trend is confirmed when we look at the contamination of Galactic dust in the reconstructed TSZ mean profile (Fig.  \ref{Fig:sampledustflux}): residual Galactic dust has a larger amplitude in the ACT-like reconstruction. When we are interested in reconstructing the TSZ profiles of compact clusters with $\theta_{200} < 5~ \mathrm{arcmin}$ (middle panels and bottom panels of Fig. \ref{Fig:sampleflux}) we can take advantage of the high resolution ACT-like channel-maps in the NILC combination to go beyond the limited beam of the Planck-like instrument. The middle left and bottom left panels of Fig. \ref{Fig:sampleflux} confirm that a component separation applied on the single Planck-like dataset fails to recover the TSZ effect from compact clusters. 

\begin{table}
\begin{flushleft}
\begin{tabular}{|p{2.5cm}|*{4}{c}|}
\hline
 MSE (extended)  &  Planck-like & ACT-like & Joint \\
\hline 
  & $3.95\mathrm{e}$--$11$ & $1.23\mathrm{e}$--$10$ & $3.35\mathrm{e}$--$11$ \\
\hline
\hline
 MSE (``typical'')  &  Planck-like & ACT-like & Joint \\
\hline
& $1.27\mathrm{e}$--$9$ &  $3.74\mathrm{e}$--$10$ & $8.12\mathrm{e}$--$11$\\
\hline
\hline
 MSE (compact)  &  Planck-like & ACT-like & Joint \\
\hline
& $6.52\mathrm{e}$--$7$ &  $3.65\mathrm{e}$--$10$ & $1.45\mathrm{e}$--$10$\\
\hline
\end{tabular}
\end{flushleft}
\caption{Performance of the mean sample profile reconstruction measured by the criterion Eq. (\ref{qqq:mse_sample}).}  
\label{tab:chi2_sample}
\end{table} 

For completeness we compute the mean squared error measuring the trade-off between the bias and the variance of the profile reconstruction by generalising the criterion Eq. (\ref{qqq:mse}) for a single cluster to the case of a cluster sample:
\ba
\mathrm{MSE}~ = \sum_{\mathrm{ring}~ i}\left(\overline{Y}_i-\overline{Y}^{\mathrm{(in)}}_i\right)^2+\overline{\sigma}_i^2,\label{qqq:mse_sample}
\ea
where $\overline{Y}_i={1\over n}\sum_{j=1}^n \widehat{Y}^j_i$, $\overline{Y}^{\mathrm{(in)}}_i={1\over n}\sum_{j=1}^n Y^{\mathrm{(in)}j}_i$, and ${\overline{\sigma}_i^2= {1\over n^2} \sum_{j=1}^n \left(\sigma^j_i\right)^2}$. Here $n$ is the size of the sample and the quantities indexed by $j$ refer to one cluster of the sample.
In Tab. \ref{tab:chi2_sample} we have listed the results obtained for the three sets of cluster samples (extended, ``typical'', and compact) and for the three sets of exploited data (Planck-like, ACT-like, and joint Planck-like/ACT-like). The results show that the mean TSZ profile from extended clusters ($\theta_{200} > 15$ arcmin) is reconstructed with similar accuracy for the three datasets with the ACT-like reconstruction slightly more contaminated due to the lack of channels used to clean large scale Galactic foregrounds. Conversely, results in Tab. \ref{tab:chi2_sample} confirm that Planck-like NILC completely fails in reconstructing the mean TSZ profile from more compact clusters ($\theta_{200} < 5$ arcmin), due to the limited resolution of the channels compared to ACT-like channels, whereas the joint Planck-like/ACT-like NILC significantly improves the profile reconstruction by two orders of magnitude. The best bias-variance trade-off is still accomplished by the joint Planck-like/ACT-like reconstruction of the mean TSZ profile.

\section{Discussion}\label{sec:discussion}

\subsection{Impact of the size of patch and of the number of channels}\label{subsec:bias}

\begin{figure*}
  \begin{center}
    \includegraphics[width=5cm]{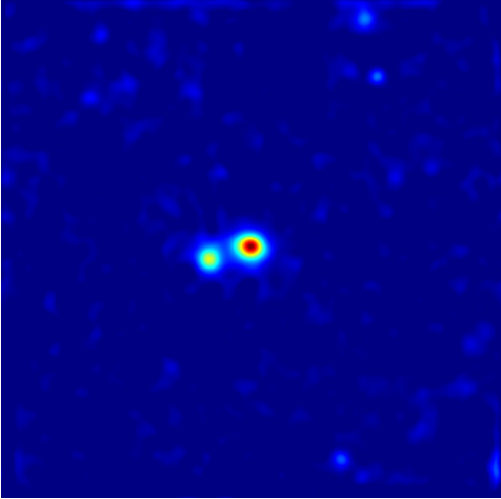}~
    \includegraphics[width=5cm]{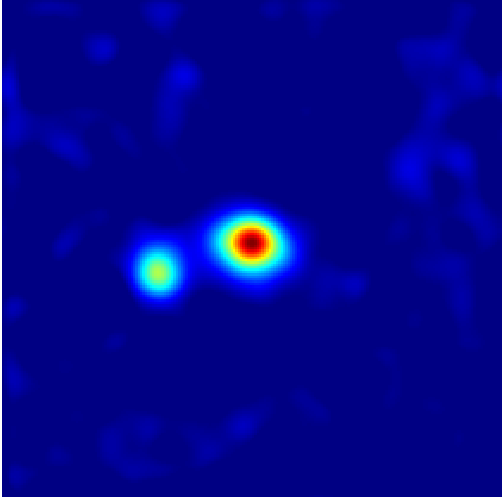}~
    \includegraphics[width=5cm]{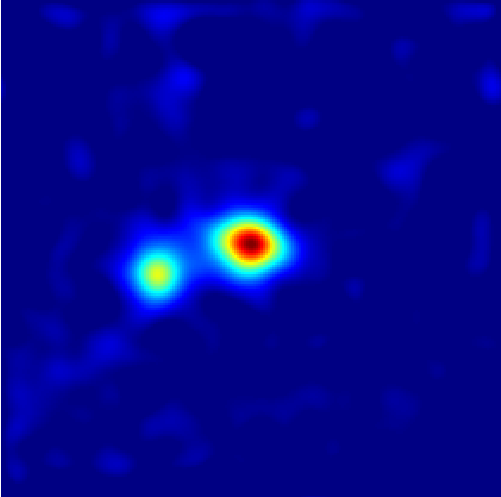}\\
     \includegraphics[width=5.3cm]{fluxall.png}
    \includegraphics[width=5.3cm]{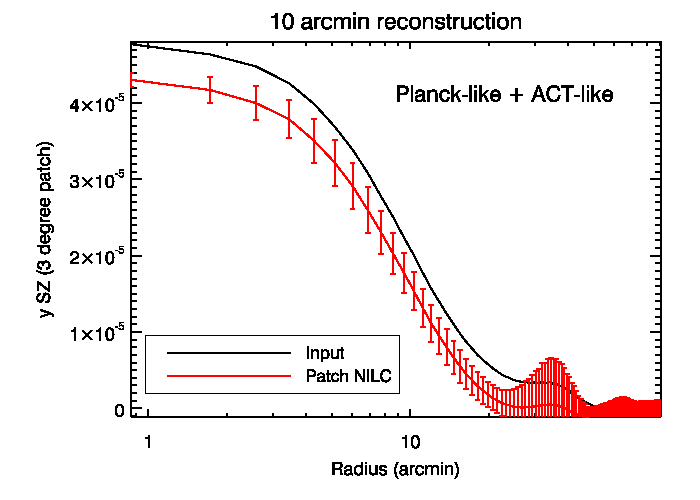}
    \includegraphics[width=5.3cm]{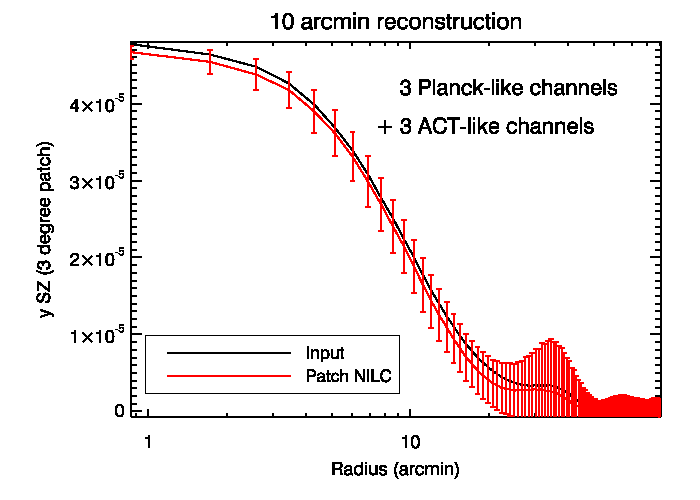}\\
     \includegraphics[width=5.5cm]{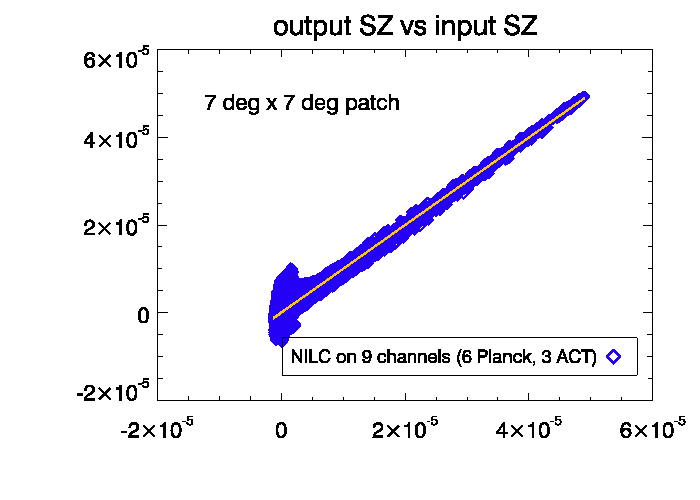}
    \includegraphics[width=5.5cm]{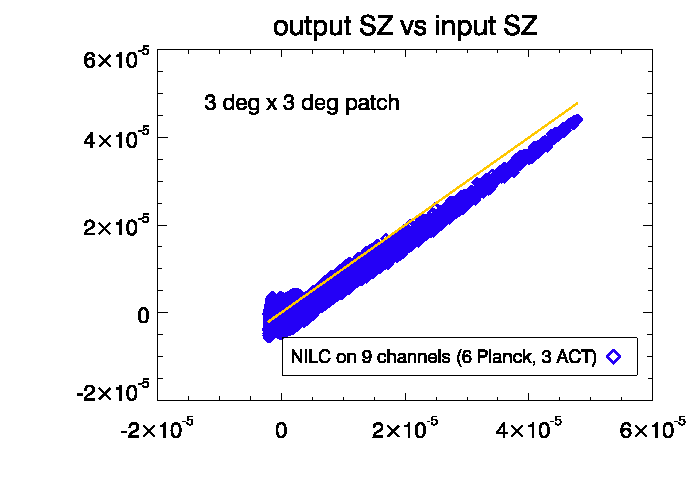}
    \includegraphics[width=5.5cm]{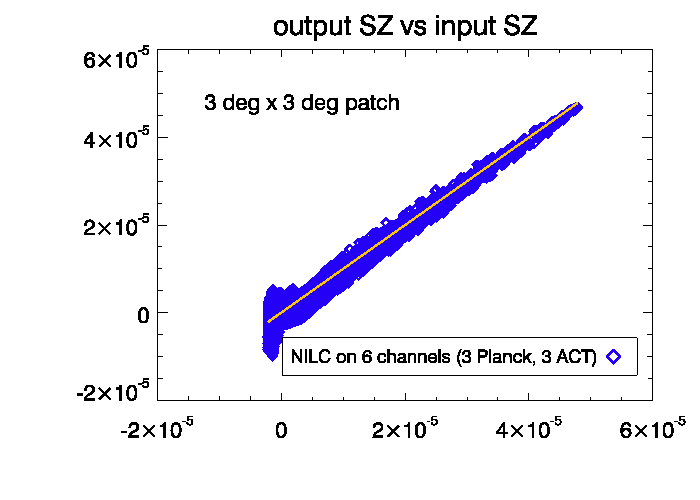}\\
  \end{center}
\caption{Impact on NILC reconstruction of the size of the patch and of the number of exploited channels. The reconstruction is shown on the selected cluster $1$ of Tab. \ref{tab:table_amas} and is performed at a resolution of $10$ arcmin. Left panels: the patch is $7^\circ\times 7^\circ$ and all Planck-like and ACT-like channels have been exploited by NILC. Middle panels: the patch is $3^\circ\times 3^\circ$ and all Planck-like and ACT-like channels have been exploited by NILC. Here a typical ILC bias is observed due to the decreasing of the number of modes in such a small patch of the sky. Right panels: in the case of a $3^\circ\times 3^\circ$ patch the bias is removed by decreasing the number of exploited channels to six channels: $3$ Planck-like channels ($30$ GHz, $90$ GHz, and $350$) and $3$ ACT-like channels ($148$ GHz, $219$ GHz, and $277$ GHz).}
\label{Fig:smallpatch}
\end{figure*}

By contrast with a full-sky ILC analysis, when performing component separation on small patches of the sky we encounter the problem of ILC bias discussed in \citet{2009A&A...493..835D}. 
 On small patches the statistics is computed on a reduced number $N_{\mathrm{p}}$ of modes\footnote{The number of independent modes corresponds to the number of needlet coefficients at a given scale (set of neighbouring pixels) used to compute the local covariance matrix Eq. (\ref{qqq:localcovar}).} so that it creates artificial correlations where the ensemble average would vanish. The artificial correlations manifest themselves as a bias in the variance of the ILC estimate $\widehat{s}$:
\ba\label{qqq:varbias}
\langle \widehat{s}^2 \rangle = \langle s^2 \rangle + \langle \left(\widehat{s}-s\right)^2 \rangle + 2\langle s\left(\widehat{s}-s\right) \rangle,
\ea
where the third term in the right-hand side of Eq. (\ref{qqq:varbias}) would vanish in the absence of bias, given the assumption that the sky components are physically independent. At first order in ${1/N_{\mathrm{p}}}$, this bias is an anticorrelation between the signal and the residual noise and is given by the simple formula:
\ba\label{qqq:bias}
\langle s\left(\widehat{s}-s\right) \rangle = -\langle s^2 \rangle {N_{\mathrm{ch}}-1 \over N_{\mathrm{p}}}
\ea
where $N_{\mathrm{ch}}$ is the number of channels and $N_{\mathrm{p}}$ is the number of modes used to compute the coefficients of the local covariance matrix Eq. (\ref{qqq:localcovar}). This unphysical anticorrelation may thus induce a power loss in the TSZ signal reconstruction. The exact derivation of formula Eq. (\ref{qqq:bias}) can be found in the appendix of \citet{2009A&A...493..835D}. It can be intuitively understood as follows. On the one hand the variance of each mode of the true signal is $\langle s^2 \rangle/N_{\mathrm{p}}$ since the sum over the modes must have total variance $\langle s^2 \rangle$. On the other hand, the number of modes available for reconstructing the TSZ signal is $(N_{\mathrm{p}}-(N_{\mathrm{ch}}-1))$ because $(N_{\mathrm{ch}}-1)$ degrees of freedom are required to adjust the $N_{\mathrm{ch}}$ weights $w_i$ of the constrained minimization problem Eq. (\ref{qqq:opt}). 
 Therefore, the variance of each mode of the reconstructed signal by ILC is $\langle s^2 \rangle/(N_{\mathrm{p}} - (N_{\mathrm{ch}}-1))$. The
correlation of the reconstructed TSZ map with the original TSZ map thus is $(N_{\mathrm{p}} - (N_{\mathrm{ch}} - 1))/N_{\mathrm{p}}$. It is this loss of $N_{\mathrm{ch}}  - 1$ modes of the original TSZ which induces the negative bias  Eq. (\ref{qqq:bias}).

On infinite samples or large patches, $N_{\mathrm{p}}$ goes to infinity so that the bias Eq. (\ref{qqq:bias}) goes to zero. Conversely, decreasing the size of the patches results in the increase of the bias. This is shown in Fig.  \ref{Fig:smallpatch} (middle versus left panels) where we compare the NILC TSZ reconstruction on a $3^\circ\times 3^\circ$ patch with the reconstruction on a $7^\circ\times 7^\circ$ patch centred at the same location on the sky. 
 In order to reduce the bias on small patches we should reduce the number of channels used in the ILC filtering, according to formula Eq. (\ref{qqq:bias}). However, the number of channels has to be significant enough in order to enable ILC to correctly filter the contamination from foregrounds. Therefore, we are left with a compromise between minimizing the signal bias versus minimizing the background noise. On the right panels of Fig.  \ref{Fig:smallpatch} we show that we are able to remove the ILC bias on a $3^\circ\times 3^\circ$ patch TSZ reconstruction by reducing the number of exploited channels to six ($30$ GHz, $90$ GHz, and $350$ GHz Planck-like channels, and the three ACT-like channels) while this number of channels is kept significant to minimize the background noise from the other sky components. We deliberately selected a subset of six channels covering the whole range of available frequencies going from $30$ GHz to $350$ GHz in order to guarantee a robust cleaning of the residuals from other sky emissions in the NILC TSZ reconstruction.



\begin{figure*}
  \begin{center}
    \includegraphics[width=5.5cm]{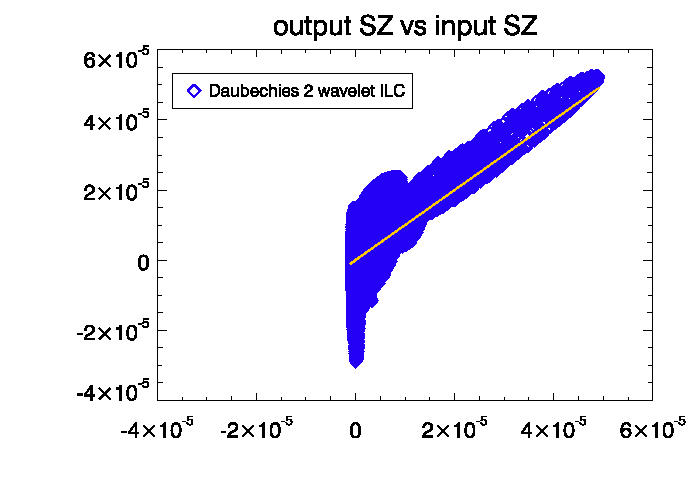}
    \includegraphics[width=5.5cm]{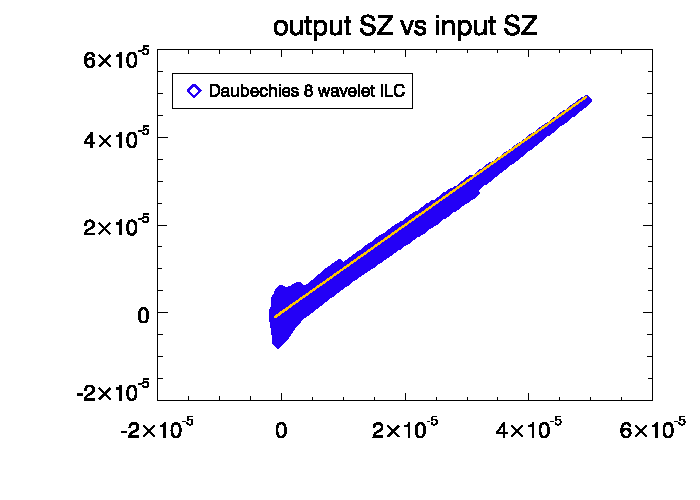}
    \includegraphics[width=5.5cm]{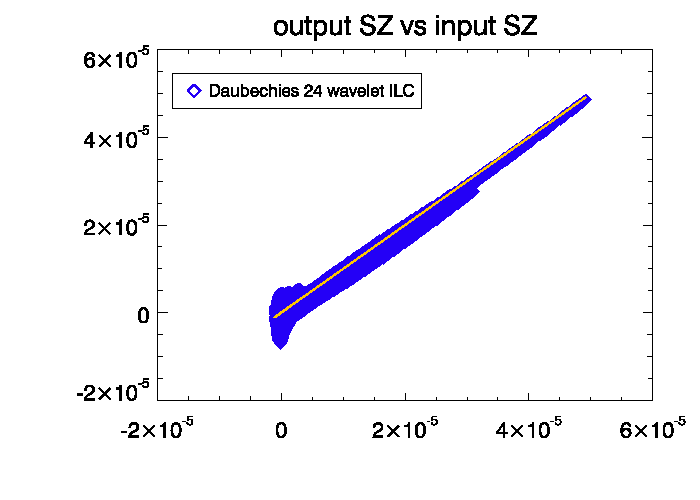}
    \includegraphics[width=5.5cm]{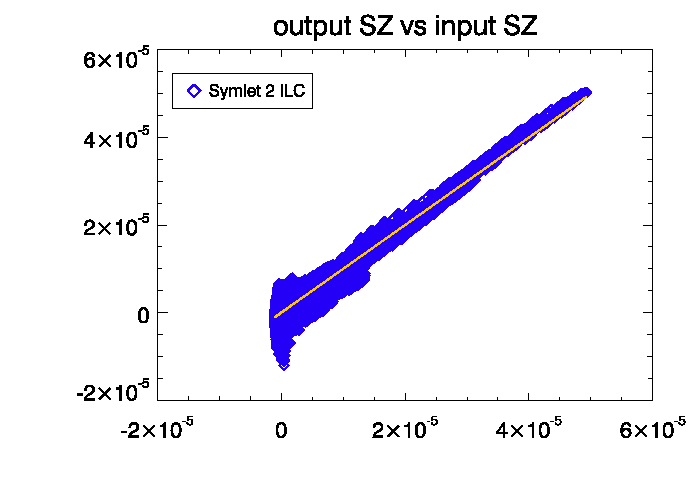}   
   \includegraphics[width=5.5cm]{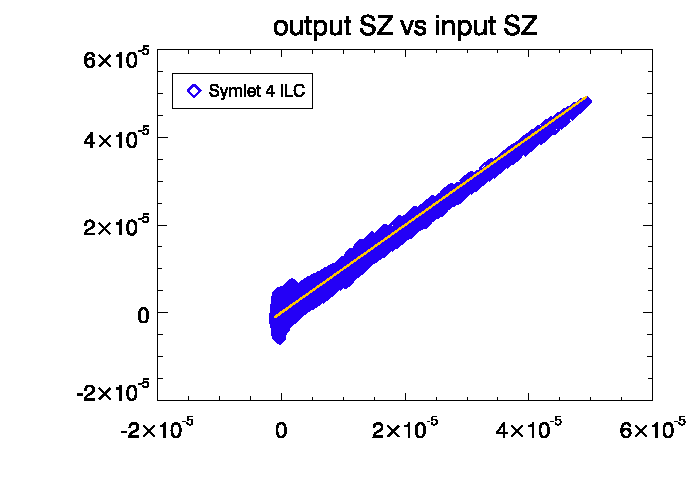}
   \includegraphics[width=5.5cm]{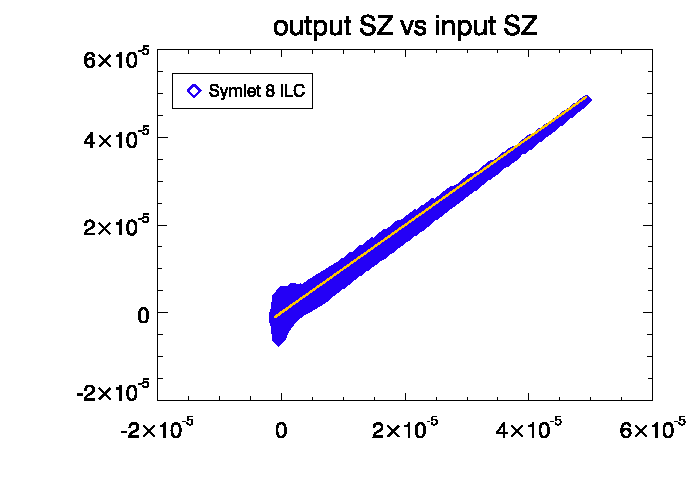}
    \includegraphics[width=5.5cm]{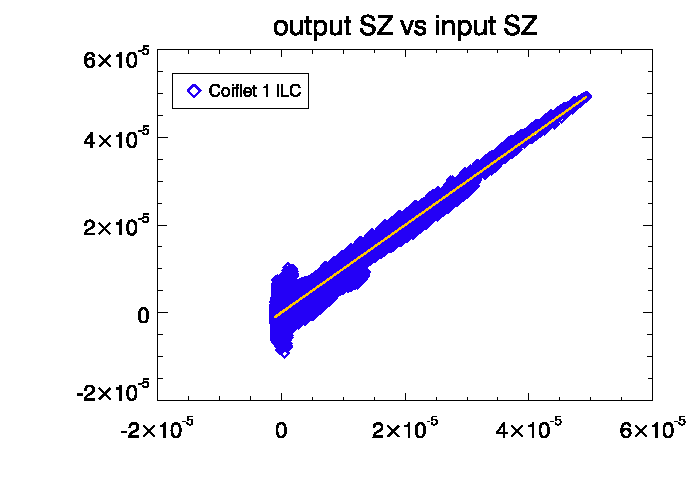}
    \includegraphics[width=5.5cm]{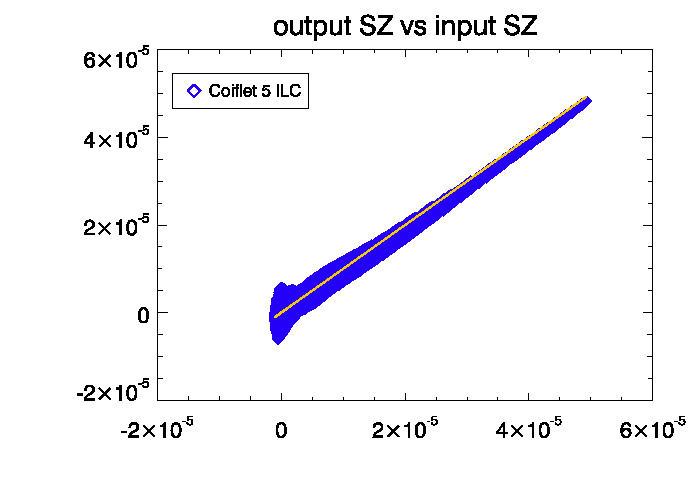}
    \includegraphics[width=5.5cm]{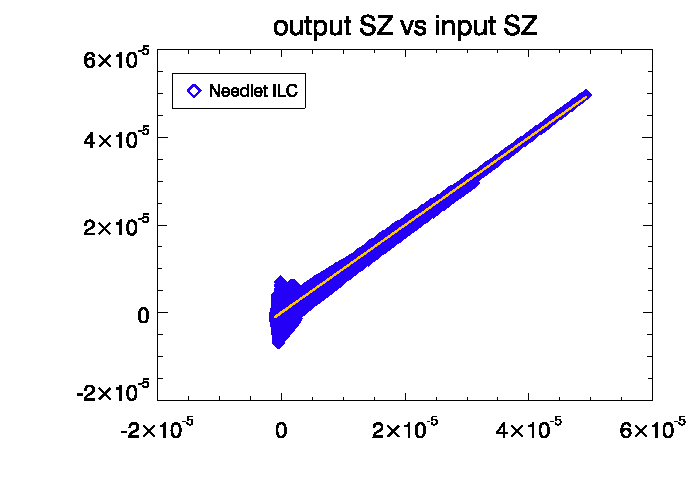}
  \end{center}
\caption{Wavelet ILC comparison (scatter plots) for the selected cluster $1$ of Tab. \ref{tab:table_amas}. \underline{TOP}: Daubechies wavelets of order 2, 8, and 24. \underline{MIDDLE}: Symlets of order 2, 4, and 8. \underline{BOTTOM}: Coiflets of order 1, 5, and Needlets. } 
\label{Fig:wavcomp}
\end{figure*}

\begin{figure*}
  \begin{center}
    \includegraphics[width=5.5cm]{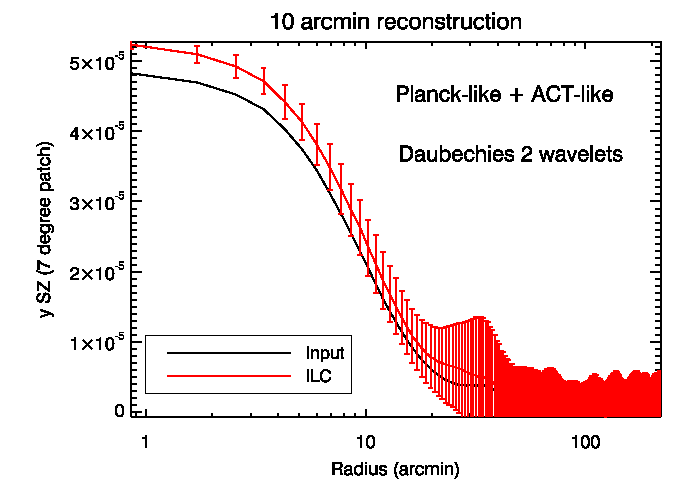}
    \includegraphics[width=5.5cm]{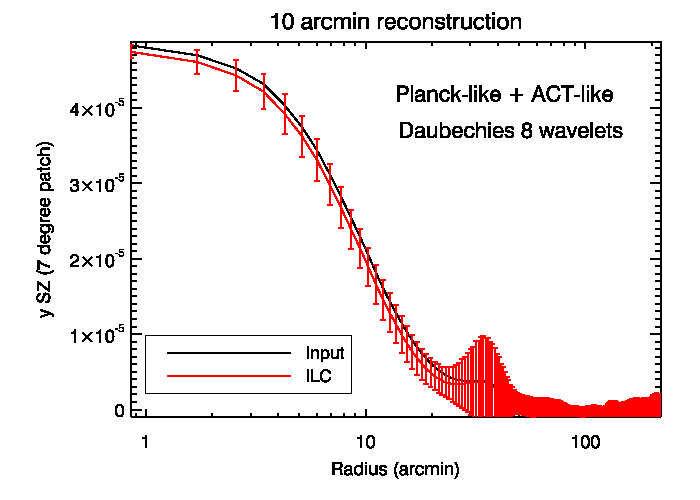}
    \includegraphics[width=5.5cm]{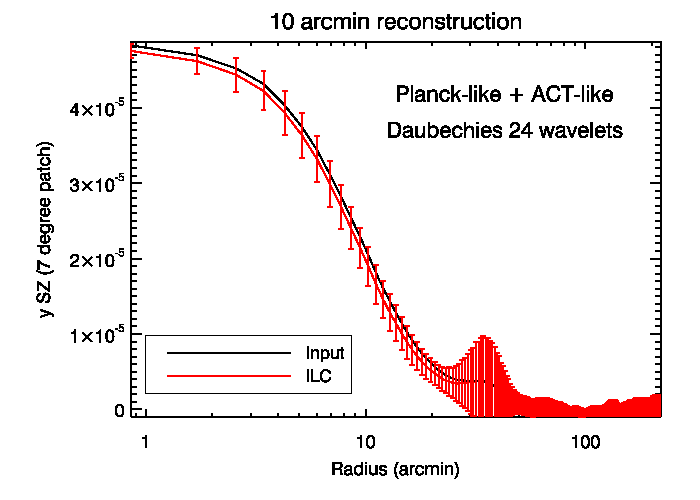}
    \includegraphics[width=5.5cm]{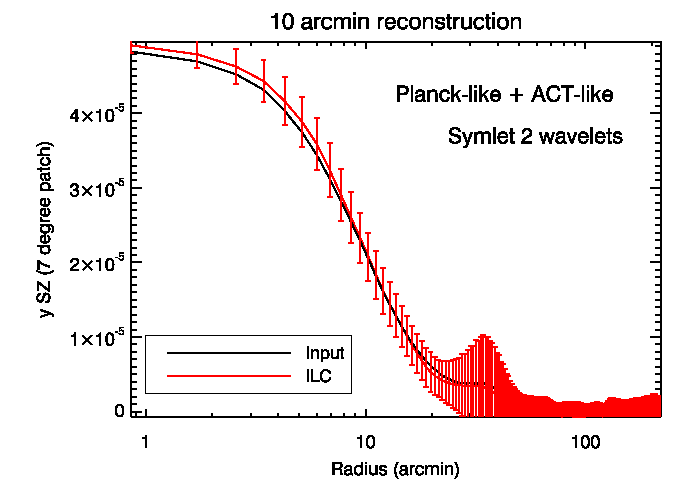}   
   \includegraphics[width=5.5cm]{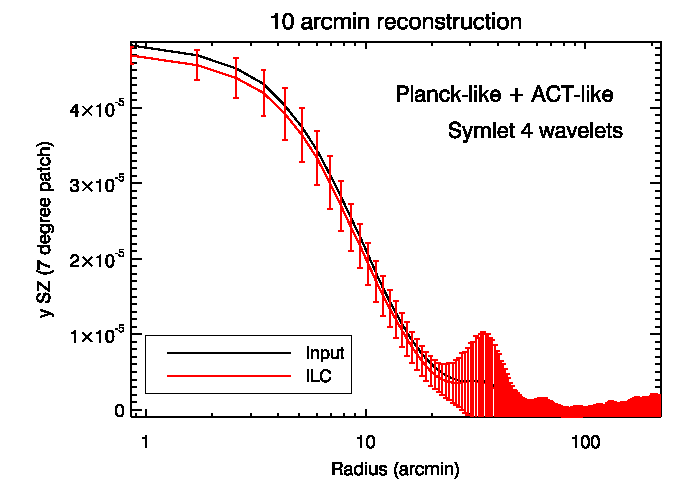}
   \includegraphics[width=5.5cm]{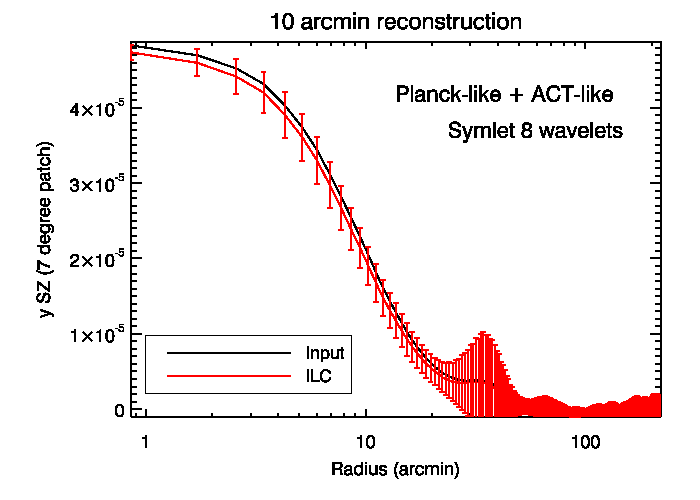}
    \includegraphics[width=5.5cm]{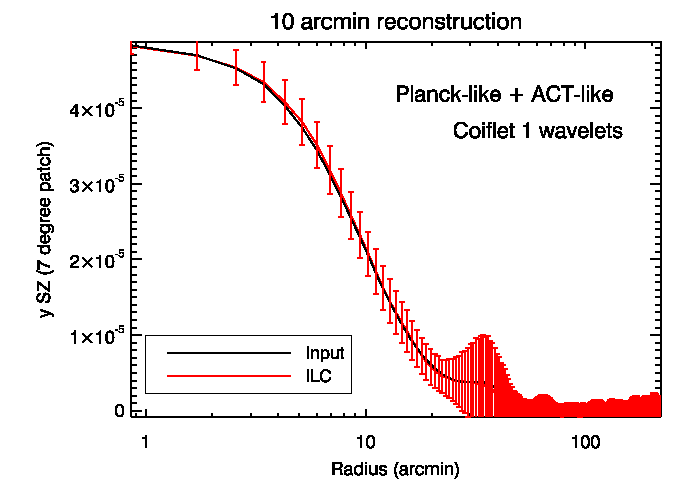}
    \includegraphics[width=5.5cm]{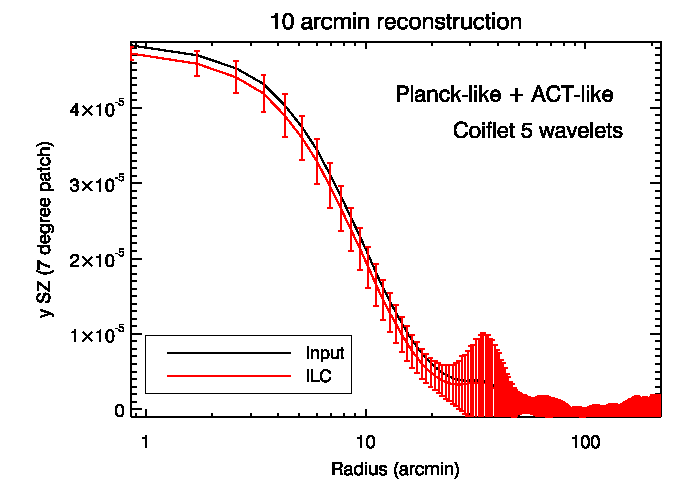}
    \includegraphics[width=5.5cm]{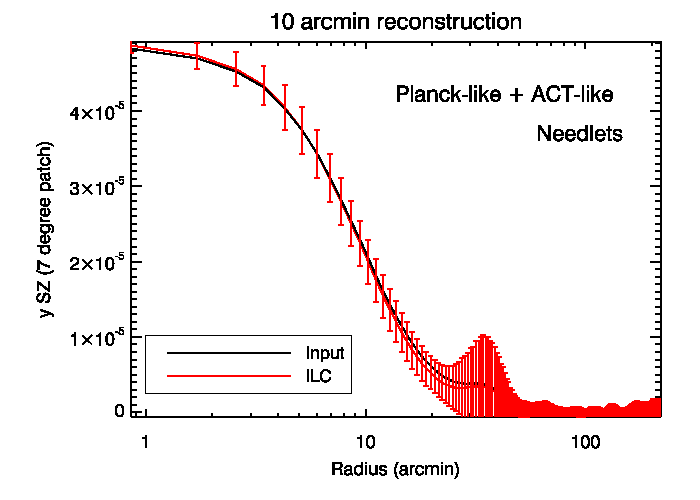}
  \end{center}
\caption{Wavelet ILC comparison (TSZ profile) for the selected cluster $1$ of Tab. \ref{tab:table_amas}. \underline{TOP}: Daubechies wavelets of order 2, 8, and 24. \underline{MIDDLE}: Symlets of order 2, 4, and 8. \underline{BOTTOM}: Coiflets of order 1, 5, and Needlets. } 
\label{Fig:wavcompflux}
\end{figure*}

\subsection{Type of wavelets}


We have shown in Sect. \ref{subsec:example} that needlets enables ILC to combine  heterogeneous instrument datasets with different resolutions for TSZ reconstruction.
 For comparison, we repeat TSZ component separation combining both Planck-like and ACT-like observations with an ILC based on different types of wavelets: Daubechies wavelets, symlets, coiflets (see \citet{1992tlw..conf.....D} and the references therein) and needlets \citep{guilloux:fay:cardoso:2008}. The definition of the standard wavelets (Daubechies, symlets, coiflets) is such that the dilation factor is sampled on a dyadic scale. The results on the reconstruction of the TSZ signal from the selected cluster $1$ of Tab. 
\ref{tab:table_amas} are shown in Figs. \ref{Fig:wavcomp} and \ref{Fig:wavcompflux}.

We display in Fig. \ref{Fig:wavcomp} the scattering of the wavelet-ILC reconstructed signal with respect to the input TSZ signal. The wavelets showing the minimum residual noise are the symlets of order $8$, the coiflets of order $5$, and the needlets. Among these three types of wavelet-ILC, the needlet ILC shows a better reconstruction of the cluster profile in Fig. \ref{Fig:wavcompflux}, the other two being slightly biased. The robustness of the wavelet-ILC reconstruction depends on the properties of localization of each type of wavelet both in space and in scale. 
 The fixed dyadic sampling of the Fourier scales probed by the standard wavelet-ILC is responsible for a reduced number of modes in the reconstruction of the signal whereas a needlet ILC appears more flexible to sample the Fourier domain (number and width of needlet bands freely tunable) as discussed in \citet{guilloux:fay:cardoso:2008}. The loss of Fourier modes by a fixed sampling of the wavelets may induce some bias in the reconstruction compared to the needlet ILC reconstruction. 
From the basic tests made here, we can see in Figs. \ref{Fig:wavcomp} and \ref{Fig:wavcompflux} that the best compromise between a minimum bias on the TSZ profile and a minimum background noise in the reconstructed signal is accomplished by the use of needlets.
 A detailed analysis on wavelet localization would be required to optimize the TSZ component separation.

\section{Conclusion}\label{sec:conclusion}

The reconstruction of galaxy cluster profiles through thermal SZ effect is an exciting challenge in data analysis of current CMB experiments because the SZ profile is a powerful probe of the baryon physics in the cluster outskirts, beyond the radius of virialisation in the intra-cluster medium. In this region, X-ray measurements fail to reconstruct the thermal pressure profile. For the first time, the SZ profiles of some galaxy clusters have been reconstructed from the surveys of the high resolution telescopes ACT \citep{2011ApJ...732...44S} and SPT \citep{2010ApJ...716.1118P}. Results on the SZ profile from the Planck survey have also been published very recently in \citet{2012arXiv1207.4061P}.
Both in ACT and SPT results, the SZ signal has been reconstructed by using a matched spatial filtering, 
 which is an effective way of reconstructing the SZ effect when few frequency maps are available from the observations. However, matched filtering relies on priors on the expected template profile, which may appear controversial when one is interested in reconstructing the cluster profile. The needlet ILC on patches developed in this work is a blind approach which does not suffer from any prior to reconstruct the SZ profiles.  

Apart from the prior issue, the accuracy of the SZ reconstruction also relies on the achievable resolution of the instrument and on the level of contamination either by sky emissions (Galactic foregrounds, CMB, etc) or by the instrumental noise. We have highlighted that both problems can be tackled simultaneously by combining extra frequency maps from multiple instrument datasets with different resolutions. We have shown in this work how needlet ILC has the ability to adapt the needlet windows in Fourier space to the beams of the channel-maps exploited for component separation, thus offering the possibility of combining heterogeneous datasets coming from multiple instruments with various resolutions and different frequency coverages. On the one hand, the properties of localization of the needlets both in scale and in space enable component separation to adapt to the local conditions of contamination (physical at large scales, instrumental at small scales), on the other hand they enable component separation to aggregate heterogeneous instrument datasets with different achievable resolutions. 
 We have applied needlet ILC on three different simulated datasets (single Planck-like, single ACT-like, and joint Planck-like/ACT-like) for different patches of the sky centered on various extended and compact clusters.  The performance of the reconstruction has been validated both on particular clusters and on samples of selected clusters. The aggregation of multiple datasets with needlet ILC allows us to reconstruct the SZ profile over a large radius with both high resolution and robust foreground and noise cleaning.

We have also explored the limitations of the method. Needlet ILC applied on a small-size patch of the sky encounters a bias in the reconstructed SZ signal. However, combining a subset of channels in that case enables ILC to tackle this problem of bias. An optimized selection of the subset of channels would be required to improve the reconstruction on very small patches. It would also be very instructive to optimize in a future work the wavelet localization (type of wavelets, width of spectral windows) for multi-instrument SZ component separation. 

Needlet ILC performed on the joint dataset including both Planck-like and ACT-like datasets successfully reconstructs the TSZ effect from compact clusters (${\theta_{200} < 5~ \mathrm{arcmin}}$) beyond the beam of the single Planck-like instrument while cleaning foreground residuals better than a component separation applied on the single ACT-like instrument. 
The multiresolution approach presented in this work appears to be promising for a future multi-instrument analysis of the SZ effect.

\section*{Acknowledgements}
\thanks{We would like to thank Neelima Sehgal and Kavilan Moodley for useful correspondence. The authors acknowledge the use of the HEALPix package \citep{gorski05}. The authors also acknowledge the use of the Legacy Archive for Microwave Background Data Analysis (LAMBDA).
}

\bibliography{joint_analysis}

\label{lastpage}

\end{document}